\documentclass[aps,prd,showpacs,superscriptaddress,amsmath]{revtex4}
\usepackage{longtable}
\usepackage{graphicx}
\usepackage{times}
\usepackage{amssymb}
\usepackage{amsmath}

\newcommand{\be}{\begin{equation}}
\newcommand{\ee}{\end{equation}}
\newcommand{\beq}{\begin{equation}}
\newcommand{\eeq}{\end{equation}}
\newcommand{\bea}{\begin{eqnarray}}
\newcommand{\eea}{\end{eqnarray}}
\newcommand{\bdm}{\begin{displaymath}}
\newcommand{\edm}{\end{displaymath}}
\newcommand{\drm}{{\rm d}}

\DeclareMathOperator\e{e}
\DeclareMathOperator\irm{i}

\begin{document}
\title{BBO and the Neutron-Star-Binary Subtraction Problem}

\author{Curt Cutler}
\affiliation{Jet Propulsion Laboratory, California Institute of Technology, Pasadena, CA 91109}

\author{Jan Harms}
\affiliation{Max-Planck-Institut f\"ur Gravitationsphysik and Universit\"at Hannover, Callinstra\ss e 38, 30167 Hannover, Germany}

\date{\today}

\begin{abstract}
The Big Bang Observer (BBO) is a proposed space-based
gravitational-wave (GW) mission designed primarily to search for
an inflation-generated  GW background in the frequency range $\sim
10^{-1}\,{\rm Hz} -1\,$Hz.  The major astrophysical foreground in
this range is gravitational radiation from inspiralling compact
binaries. This foreground is expected to be much larger than the
inflation-generated background, so to accomplish its main goal,
BBO must be sensitive enough to identify and subtract out
practically all such binaries in the observable universe.   It is
somewhat subtle to decide whether BBO's current baseline design is
sufficiently  sensitive for this task, since, at least initially,
the dominant noise source impeding identification of any one
binary is confusion noise from all the others (rather than
instrumental noise). Here we present a self-consistent scheme for
deciding whether BBO's baseline design is indeed adequate for
subtracting out the binary foreground. We conclude that the
current baseline should be sufficient. 
However, if BBO's sensitivity were 
degraded by a factor $2$ from the current baseline, then its ability to 
detect an underlying primordial background would 
depend critically on the value of $\rho_{\rm th}$, the threshold
signal-to-noise ratio marking the boundary between detectable
and undetectable sources. If BBO's sensitivity were degraded by
a factor $4$ from the current baseline, it could not detect 
a primordial background below $\Omega_{\rm GW} \sim 10^{-15}$.

It is impossible to perfectly subtract out each of the binary inspiral waveforms, so an important
question is how to deal with the "residual" errors in the post-subtraction
data stream. We sketch a strategy of "projecting out"
these residual errors, at the cost of some effective bandwidth.
We also provide estimates of the sizes of various post-Newtonian effects
in the inspiral waveforms that must be accounted for in the BBO analysis.

\end{abstract}

\maketitle 

\section{Introduction}
The Big Bang Observer (BBO) is a proposed space-based
gravitational wave (GW) mission designed to search for stochastic
gravitational-wave background generated in the very early universe
\cite{Phi2003,UnEA2005}.  The design goal is to be able to detect
primordial GWs with energy density $\Omega_{\rm GW}(f) \gtrsim
10^{-17}$ in the frequency band $10^{-1}\,$Hz $ < f < 1\,$Hz.
Standard, slow-roll inflation predicts $\Omega_{\rm GW}(f)
\lesssim 10^{-16}-10^{-15}$~\cite{Tur1997}.

To achieve this sensitivity to a primordial GW background,
it will {\it first} be necessary to  subtract
from the BBO data stream the GW foreground generated by $\sim 10^5-10^6$
neutron star-neutron star (NS-NS), neutron star-black hole (NS-BH),
and black hole-black hole (BH-BH) binary mergers, out to $z \sim 5$. This
foreground "noise" has an amplitude substantially greater than BBO's
instrumental noise, which in turn is probably substantially greater
than the amplitude of the sought-for primordial GWs.
To achieve BBO's goal, the GWs from the merger foreground must
be subtracted to a level well {\it below} that of the primordial
background. This means that the amplitude of the {\it residual},
post-subtraction foreground  must be $\lesssim  10^{-2.5}$ of the pre-subtraction
level.

Will it be possible for BBO data analysts to subtract out the binary merger foreground to
this accuracy?
This question
is non-trivial to answer precisely because confusion noise from
unresolved mergers can in principle dominate the BBO noise
spectrum. To decide which mergers are unresolvable, one
needs to know the full BBO noise curve, {\it including} the level of
confusion noise from the unresolvable mergers. But to determine the
level of confusion noise, of course one needs to know which mergers are unresolvable.
Clearly, one needs somehow to solve both these
problems simultaneously.

The focus of our investigations will be on NS-NS
mergers, since these are the most problematic for BBO.
The less numerous BH-BH and BH-NS merger
events will have higher signal-to-noise ratios and therefore should be
easier to subtract. If we find the NS-NS mergers can be almost
fully subtracted from the BBO data stream, then the same should be true
for the BH-BH and BH-NS mergers.

How, in practice, will almost all the NS-NS mergers be subtracted out? We imagine that
something like the following iterative scheme could be used: begin by resolving
and subtracting out the brightest merging binaries (i.e., those with highest
signal-to-noise-ratio), then resolve and
subtract the next brightest ones, etc - regularly updating {\it all} the
parameters of the subtracted binaries, as one goes along, to give the
best global fit. Each subtraction decreases the foreground confusion
noise and so increases the distance out to which NS binaries can be
resolved. Will such a scheme suffice for BBO?  The aim of this paper is to
answer that  question {\it without actually having to carry out the
whole procedure}.
We develop a method for determining the likely efficacy of foreground
subtraction in a self-consistent manner.  Our method is (very roughly) as
follows. Imagine that BBO is surrounded by a huge sphere out to some
redshift ${\bar z}$, such that NS-NS mergers
inside the sphere (i.e., at redshifts less than ${\bar z}$) can all be
individually resolved and subtracted (using realistic computational
power), while none of the sources outside the sphere is resolvable.
This redshift $\bar z$ marking the boundary of the resolvable sources is
not known initially, so we start with a reasonable guess. We then
calculate the confusion noise due to all NS-NS mergers (NSm) at redshifts greater than
$\bar z$,
$S_{\rm h}^{{\rm NSm}, > {\bar z}}(f)$, which we add to the instrumental noise
$S_{\rm h}^{\rm inst}(f)$ to obtain the total noise:
\begin{equation}\label{eq:tot1}
S_{\rm h}^{\rm tot}(f) = S_{\rm h}^{\rm inst}(f) + S_{\rm h}^{{\rm NSm}, >{\bar z}}(f) \, .
\end{equation}
One can use this total noise level, $S_{\rm h}^{\rm tot}(f)$, to improve one's estimate
of $\bar z$, and iterate this procedure until $\bar z$ converges.

Actually, of course, the detectability of any particular NS-NS binary depends
not just on its  distance (or redshift), but also on $\mu \equiv \hat L \cdot \hat N$, where
$\hat L$ is the normal to the binary's orbital plane and $\hat N$ points along our line-of-sight.
(The binary's detectability also depends, of course,
on the other three angles describing the binary's orientation and position on the sky, but to
a much lesser extent.)
Our calculation {\it does} properly account for the $\mu$-dependence of the binary's detectability; i.e,
we take $\bar z$ to be a function of $\mu$, not a single number.

We stress that there are actually
two different sorts of confusion noise associated with
merging binaries: the full signals from unresolved binaries (mentioned above), and
the small {\it errors} that inevitably occur when waveforms from
resolved mergers are subtracted out of the data.
In Sec.~IV  we propose a method for dealing with these {\it residual}
errors, by projecting out  the subspace
in which these errors can lie, at the cost of some bandwidth.
We also estimate that this fractional decrease in BBO's bandwidth
is small enough that for our purpose (deciding whether an iterative subtraction scheme is feasible)
it can be neglected.

We remark that our calculation is quite similar in spirit to a recent analysis of
WD-binary subtraction in LISA data analysis, by Cornish et al.~\cite{CoCr2005}, which appeared
when our own work was already at an advanced stage.  In both cases, the idea is to use
the requirement of {\it self-consistency} to arrive at a unique estimate of the
efficacy of foreground subtraction,
without actually coding up the whole analysis pipeline and testing it on
simulated data.

We also remark that a recent paper by Buonanno et al.~\cite{BuEA2005} estimates
that supernova explosions {\it could} provide another important BBO foreground, via
the GW memory effect, but only if  the anisotropy of neutrino emission is quite high, on average.
For the rest of this paper we will neglect the possibility of a large foreground from supernovae.

The organization of the rest of this paper is as follows. In Sec.~II we give a brief overview of the
BBO mission, its design sensitivity, and the foreground produced by merging NS binaries.
In Sec.~III we briefly explain {\it why} the most distant NS-NS binaries are effectively a noise
source when it comes to resolving more nearby ones.  In Sec.~IV we
summarize our proposed  strategy of dealing with any residual subtraction errors by
projecting them out.
In Sec.~V we provide estimates regarding the importance of eccentricity, NS spin, and
high-order post-Newtonian (PN)  effects in correctly subtracting out the resolved mergers. Besides being
important for any future implementation of a BBO analysis pipeline, this catalog of effects is
useful in estimating the threshold signal-to-noise ratio (SNR) $\rho_{\rm th}$ required to
detect NS-NS mergers.
In Sec.~VI we take a first cut at estimating $\rho_{\rm th}$, which we
assume will be set by the then-available computational power.
Our equations for self-consistently determining the efficacy of foreground subtraction
are developed in Sec.~VII.   We solve these equations for a variety of
assumptions regarding the NS merger rate, the detection threshold $\rho_{\rm th}$, 
and BBO's instrumental noise level, and
display the solutions in Sec.~VIII.
We summarize our conclusions in Sec.~IX.
The derivation of one of the equations in Sec.~VII is relegated to Appendix A.

We use units in which $G=c=1$.  Therefore, everything can be measured in
the fundamental unit of seconds. However, for the sake of familiarity,
we also sometimes express quantities in terms of yr, Mpc, or $M_\odot$,
which are related to our fundamental unit by 1 yr $= 3.1556 \times 10^7$s,
1 Mpc $= 1.029 \times 10^{14}$s, and $1 M_\odot = 4.926 \times 10^{-6}$s.

For concreteness, we assume the universe corresponds to
a flat Friedmann-Robertson-Walker model, with the universe's
matter and vacuum energy densities being given by
$\Omega_{\rm m}=0.33 $ and $\Omega_\Lambda=0.67$, respectively.
Our fiducial value for the Hubble constant is
$H_0=70\,\rm km\,s^{-1}\,Mpc^{-1}$. 

\section{Overview of BBO and the NS-binary background}

\subsection{BBO}
BBO is essentially a follow-on mission to LISA, the planned Laser Interferometer
Space Antenna~\cite{LISA2000}, but optimized to
detect GWs generated by parametric amplification during inflation.
(For a review of inflation-generated GWs, see Allen \cite{All1996} and references therein.)
In the LISA band, $10^{-5}\,$Hz -- $10^{-1}\,$Hz, an inflation-generated signal
with $\Omega_{\rm GW} \alt 10^{-15}$ would be completely covered up by
the foreground produced by galactic and extra-galactic white-dwarf binaries.
By contrast, BBO will have its best sensitivity in the range $\sim 0.1\,$Hz -- $1\,$Hz.
This band avoids the GW
foreground produced by all the white dwarf binaries in the universe,
which cuts off at $f \lesssim 0.2$ Hz (where the most
massive of the WD binaries merge).
In the BBO band,
the dominant foreground GW sources are inspiralling NS-NS, NS-BH, and BH-BH binaries.
BBO's baseline design, and corresponding instrumental noise curve, have been set
in large part by the
requirement that one must be able to individually identify
practically {\it all} such
inspiral signals and subtract them out of the data.
An initial rough estimate suggested that the baseline "specs"  in Table I are
adequate for this purpose \cite{Phi2003};  our primary task in this paper is to examine that
issue much more carefully.

\begin{figure}[ht!]
\centerline{\includegraphics[width=7.0cm]{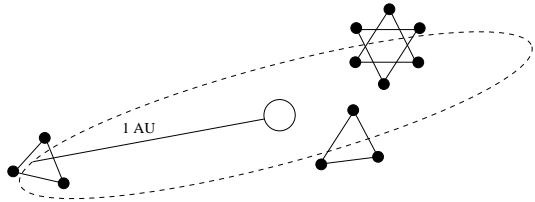}}
\vspace{0mm}
\caption[The BBO configuration]{The Big-Bang Observer (BBO) consists
of four LISA-like triangular constellations orbiting the Sun at
$1\,$AU. The GW background is measured by cross-correlating the
outputs of the two overlapping constellations.}
\label{figBBO}
\end{figure}

The current BBO design calls for four constellations of three
satellites each, all following heliocentric orbits at a distance of
1 AU from the Sun (see Fig.~\ref{figBBO}).
Each 3-satellite constellation can be thought of as a
``short-armed LISA''.
Two of the constellations
overlap to form a ``Jewish star''; the other two are ahead and behind
by $2\pi/3$ radians, respectively.
Briefly, the idea behind this orbital geometry is
that $\Omega_{\rm GW}(f)$ will be measured by cross-correlating the
outputs of the two overlapping constellations in the Jewish star (much
as LIGO attempts to measure $\Omega_{\rm GW}(f)$ by cross-correlating
the outputs of the Livingston and Hanford
interferometers~\cite{AlRo1999}). The other two constellations give
BBO its angular resolution: $\Delta \theta \sim 10^{-2}({\rm
SNR})^{-1}\,$ radians. It is not clear whether this angular resolution is
strictly necessary for the purpose of measuring $\Omega_{\rm GW}(f)$,
but it will be immensely useful for BBO's secondary goal -- to identify,
map, and accurately determine the physical parameters of practically all merging
compact binaries in the observable universe.

>From the output of each 3-satellite constellation
(i.e., each "mini-LISA"), using time-delay interferometry (TDI) one can synthesize
data streams  that are free of laser phase noise and optical bench noise~\cite{ETA2000,PrEA2002,KTV2004}.
A particularly convenient set of TDI variables to work with is 
\{$A$, $E$, $T$\}; 
all the GW information registered by each mini-LISA is 
encoded in these variables, plus the noises in 
these $3$ channels are uncorrelated with each other (i.e., they are statistically independent).
Then, for instance, it is straightforward to find, for any source, the particular combination
of \{$A$, $E$, $T$\} that yields the optimum detection statistic, and so to determine
LISA's optimum sensitivity to that source~\cite{PrEA2002}. 

For our purposes, however, the following simplified treatment is adequate.
As is clear from Fig.~4 of Prince et al.~\cite{PrEA2002},
for NS-NS inspirals, each mini-LISA's sensitivity 
(using the optimum combination of the $A,E$ and $T$ channels) is  
practically equivalent to the sensitivity of two synthetic Michelson detectors, represented
by the TDI variables $X$ and $Y$.
For our purposes, then, we can regard BBO, which is made up of 4 mini-LISAs, as formally equivalent to 
8 synthetic Michelson interferometers.

To construct the instrumental noise curve, $S^{\rm inst}_{\rm h}(f)$, of {\it each} of these synthetic Michelson's, we
used Larson's on-line ``Sensitivity curve generator''~\cite{larson_online}, plugging in 
the parameters appropriate to BBO, which are 
listed here in Table~\ref{tabParaBBO}. 

\begin{table}[ht!]
\begin{center}
\begin{tabular}{|l|c|r|}
\hline
& Symbol & Value $\quad$\\
\hline\hline
Laser power & $P$ & $300\,$W \\
Mirror diameter & $D$ & $3.5\,$m \\
Optical efficiency & $\epsilon$ & $0.3$ \\
Arm length & $L$ & $5\cdot 10^7\,$m \\
Wavelength of laser light &$\lambda$ & $0.5\,\rm{\mu}m$ \\
Acceleration noise & $\sqrt{S_{\rm acc}}$ & $3\cdot 10^{-17}\,{\rm m}/({\rm s}^2 \sqrt{{\rm Hz}})$\\
\hline
\end{tabular}
\caption{BBO parameters.} 
\label{tabParaBBO}
\end{center}
\end{table}

\begin{figure}[ht!]
\hspace*{-0.5cm}\includegraphics[width=10.0cm]{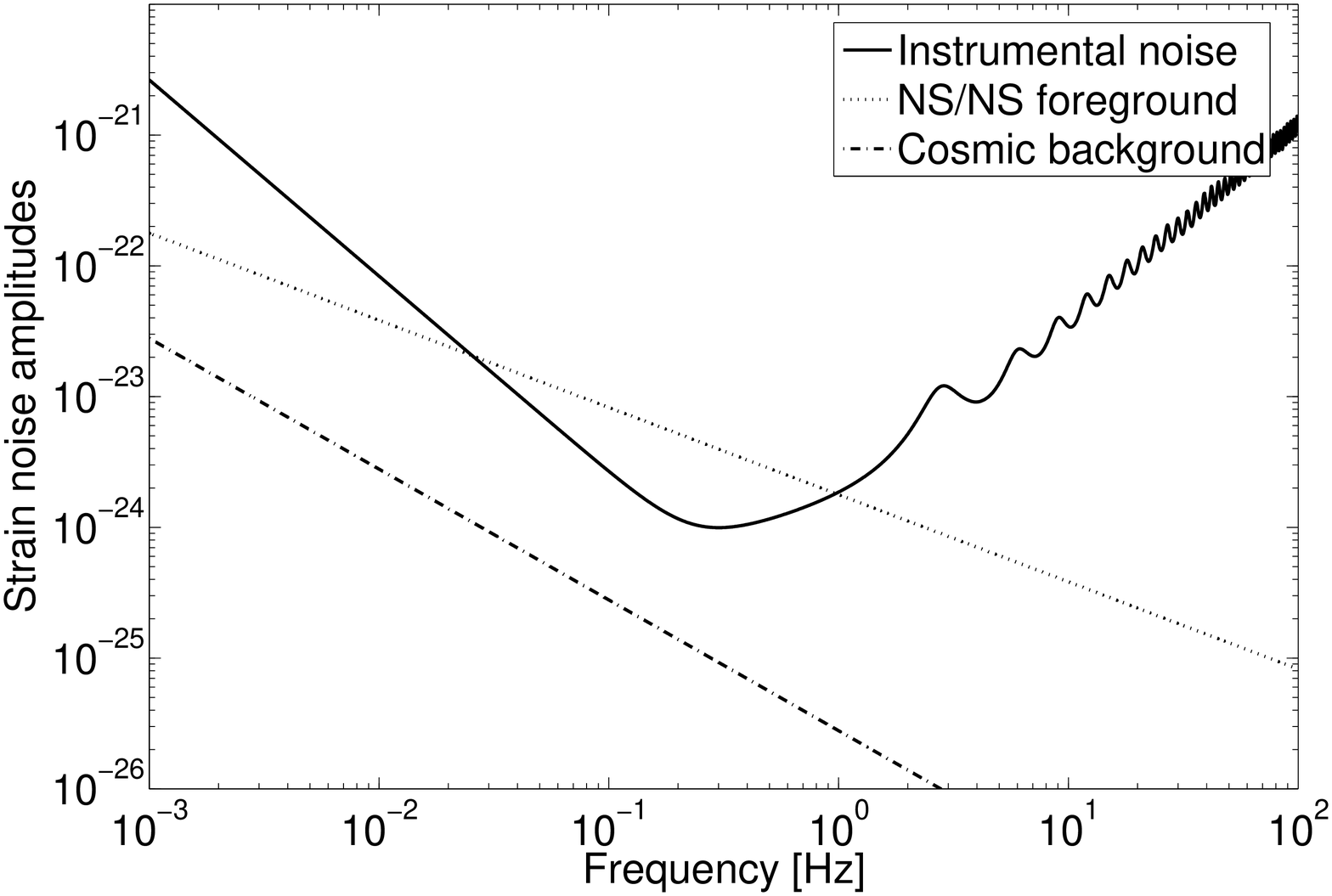}
\caption[BBO sensitivity curve]{Shows the amplitude of the
instrumental noise, $\sqrt{f S_{\rm h}^{\rm inst}(f)}$,
compared to the
amplitude of the (pre-subtraction) NS binary foreground (plotted for
$\dot n_0 = 10^{-7}\,{\rm Mpc}^{-3}{\rm yr}^{-1}$) and the sought-for
cosmic GW background (plotted for $\Omega_{\rm GW}(f) = 10^{-15}$). Clearly, to reveal
a cosmic GW background at this level, the NS foreground must be
subtracted off, with fractional residual of $\lesssim 10^{-2.5}$. }
\label{figNoiseBBO}
\end{figure}

The parameters we adopt as reference values here 
are taken from the BBO proposal~\cite{Phi2003}; these parameters do not necessarily represent the 
latest thoughts on the mission's design (which is a moving target), but do provide 
a convenient baseline for comparison.
(Reference~\cite{Phi2003} also lists parameters for less and more ambitious versions
of the BBO mission, referred to as ``BBO-lite'' and ``BBO-grand'', respectively, but in this paper
we concentrate on the intermediate version, or ``standard BBO''.)
In using the on-line generator, we have specified that the high-frequency part
of $S^{\rm inst}_{\rm h}$ is 4 times larger than the contribution from photon shot noise alone. This is the
same choice made in Fig.~1 of the BBO proposal~\cite{Phi2003}, and is consistent with assumptions typically made
in drawing the LISA noise curve. As is conventional in the LISA literature, we take $S_{\rm h}(f)$ to be
the {\it single-sided, sky-averaged} noise spectrum for each synthetic Michelson.
This BBO instrumental noise curve is shown in Fig.~\ref{figNoiseBBO}.

\subsection{The NS-NS merger rate and the associated foreground noise level}
In this subsection we estimate the magnitude of the
GW foreground from all NS-NS mergers.
We denote the NS-NS merger rate (per unit proper time, per unit
co-moving volume) at redshift $z$ by $\dot n (z)$.
The present-day density $n_0$ of merger remnants is related to $\dot n(z)$ by~\cite{Phi2001}
\beq
n_0  =\int\limits_0^\infty\drm z\frac{\dot{n}(z)}{(1+z)H(z)} \, ,
\eeq
where
\beq\label{Hz}
H(z)\equiv H_0 \sqrt{\Omega_{\rm m}(1+z)^3 + \Omega_\Lambda} \, .
\eeq
As is conventional, we define $\Omega_{\rm GW}(f)$ to be the universe's fractional energy in
GWs, per logarithmic frequency interval:
\beq
\Omega_{\rm GW}(f) \equiv \frac{1}{\rho_{\rm c}}\frac{\drm \rho_{\rm GW}(f)}{\drm(\ln\,f)} \, ,
\eeq
where $\rho_c = 3 H^2_0/(8\pi)$ is  the universe's current energy density.
Then the GW energy density (in the BBO band) due to (the inspiral phase of) all NS-NS mergers is given by \cite{Phi2001}
\bea\label{eq:omega}
\Omega_{\rm GW}^{\rm NSm}(f) & = &
\frac{8 \pi^{5/3}}{9} \frac{1}{H_0^2} {\cal M}^{5/3} f^{2/3} n_0 \langle(1+z)^{-1/3}\rangle \nonumber \\
&=& 1.7 \times 10^{-12} h_{70}^{-2}
\bigg(\frac{{\cal M}}{1.22 M_{\odot}}\bigg)^{5/3}\bigg(\frac{f}{1\,{\rm Hz}}\bigg)^{2/3} \nonumber \\
&&\qquad \cdot\bigg(\frac{n_0}{10^3\,{\rm Mpc}^{-3}}\bigg)\bigg(\frac{\langle(1+z)^{-1/3}\rangle}{0.80}\bigg) 
\eea
The term $\langle(1+z)^{-1/3}\rangle$ in Eq.~(\ref{eq:omega}) is the
merger-rate-weighted average of $(1+z)^{-1/3}$, given by
\beq
\langle(1+z)^{-1/3}\rangle\equiv\frac{1}{n_0}\int\limits_0^\infty\drm z\frac{\dot{n}(z)}{(1+z)^{4/3}H(z)}.
\label{eqNSfracD}
\eeq
What is the universe's NS-NS merger rate history, $\dot n(z)$?
It is convenient to regard $\dot n(z)$ as the product of two factors:
\beq
\dot n(z) = \dot{n}_0\cdot r(z) \, ,
\eeq
where $\dot{n}_0$ is the merger rate today and $r(z)$
encapsulates the rate's time-evolution.

For $r(z)$, we adopt the following
piece-wise linear fit to the rate evolution derived in \cite{SFMPZ2001}:
\beq\label{rz}
r(z) = \begin{cases}1+2z & z\leq 1\\
\frac{3}{4}(5-z) & 1\leq z\leq 5 \\ 0 & z\geq 5\end{cases}
\eeq
For this $r(z)$ and our fiducial cosmological model, one has
\beq
n_0 = \dot n_0 \cdot \left(2.3\cdot 10^{10}\,{\rm yr}\right) ,
\eeq
and $\langle(1+z)^{-1/3}\rangle = 0.82$.
(We note  that, as stressed in~\cite{Phi2001},
the value of $\langle(1+z)^{-1/3}\rangle$ is actually quite insensitive to
one's choice of the function $r(z)$, generally being in the range
$\sim 0.7-0.9$.)

The current NS-NS merger rate, $\dot{n}_0$, is also usefully
regarded as the product of two factors:
the current
merger rate in the Milky Way and a factor that extrapolates from the
Milky Way rate to the average rate in the universe. The NS-NS merger
rate in the Milky Way has been estimated by several authors; it is
still highly uncertain, but most estimates are in the range
$10^{-6}\,-10^{-4}\,{\rm yr}^{-1}$ \cite{BKB2002,KNST2001,VoTa2003}.
To extrapolate to the rest of the
universe, Kalogera et al.~\cite{KNST2001} estimate that one should
multiply the Milky Way rate by 
$1.1-1.6 \times 10^{-2}\cdot h_{70}^{-1}\,{\rm
Mpc}^{-3}$. That factor is obtained by extrapolating from the
B-band luminosity density
of the universe, and it is only a little larger than the extrapolation
factor derived by Phinney in \cite{Phi1991}.
Given the large overall uncertainty, in this paper we will
consider 3 possible rates: $\dot{n}_0 = 10^{-8}, 10^{-7}$, and
$10^{-6}\,{\rm Mpc}^{-3}{\rm yr}^{-1}$.

How many NS-NS merger events $\Delta N_{\rm m}$  enter the BBO band during some observation time
$\Delta\tau_0$?
Summing the contributions from all redshifts, the rate $\dot N \equiv \Delta N_{\rm m}/\Delta\tau_0$
is easily shown to be
\beq
\dot N = \int_0^{\infty} 4\pi [a_0 r_1(z)]^2 \, \dot n(z)\, \frac{\drm\tau_1}{\drm z}\, \drm z \, ,
\eeq
where (for our fiducial cosmology)
\bea
a_0 r_1(z) & = & \frac{1}{H_0} \int_0^z \frac{\drm z'}{\sqrt{(1 - \Omega_{\Lambda})(1 + z')^3  \ + \ \Omega_{\Lambda}}} \\
\frac{\drm \tau_1}{\drm z} & = & \frac{1}{H_0}\frac{1}{1+z}\frac{1}{\sqrt{(1 - \Omega_{\Lambda})(1 + z)^3  \ + \ \Omega_{\Lambda}}}\, .
\eea
This yields
\beq
\Delta N_{\rm m} = 3.0\cdot 10^5 \left(\frac{\Delta \tau_0}{3\,{\rm yr}}\right)\left(\frac{\dot n_0}{10^{-7}\,{\rm Mpc}^{-3}{\rm yr}^{-1}}\right)
\eeq
The time required for a NS-NS inspiral signal to sweep through the BBO band
will typically be comparable to BBO's lifetime. More specifically,
the time remaining
until merger, from the moment the GW frequency sweeps through $f$, is
given (to lowest post-Newtonian order) by
\beq
t(f) = 4.64 \times
10^5\,{\rm s} \left(\frac{\mathcal{M}(1+z)}{1.22
M_{\odot}}\right)^{-5/3} \left(\frac{f}{1 {\rm Hz}}\right)^{-8/3}
\eeq
where $\mathcal{M}\equiv \mu^{3/5}M^{2/5}$ is the
so-called ``chirp mass'' of the binary. (Here $M$ is the binary's
total mass and $\mu$ is its reduced mass.) Therefore, for two $1.4
M_{\odot}$ NSs, $f \approx 0.205\,{\rm Hz},\,0.136\,{\rm
Hz},\,$ and $0.112\,{\rm Hz}$ at one year, three years, and five years before
merger, respectively.

\begin{figure}[ht]
\centerline{\includegraphics[width=9.0cm]{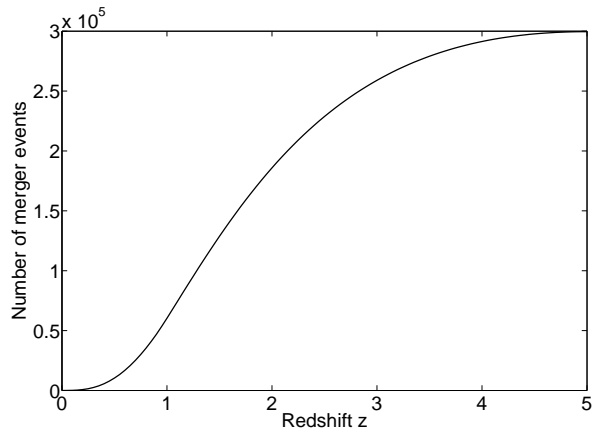}}
\caption[Total number of merger events]{
The total number of NS-NS mergers closer than redshift $z$,
The results here are normalized to a 3-yr observation period
and $\dot n_0= 10^{-7} {\rm Mpc}^{-3}{\rm yr}^{-1}$.}
\label{figNSNSnum}
\end{figure}

Figure \ref{figNSNSnum} plots the number of observable mergers during 3 years that occur closer than (any given)
redshift $z$. We see that only $\sim 15\%$ of mergers occur closer to us than $z = 1$.

The (single-sided, sky-averaged) noise spectral density associated with
any given GW background is ~\cite{All1996}:
\beq
S^{\rm GW}_{\rm h} =
\frac{3}{2}\frac{H_0^2}{\pi^2}\frac{1}{f^3}\Omega_{\rm GW}(f)
\eeq
or
\beq \big[f S^{\rm GW}_{\rm h} (f)\big]^{1/2} = 8.8 \times 10^{-25}\cdot h_{70}
\left(\frac{\Omega_{\rm GW}(f)}{10^{-12}}\right)^{1/2} \left(\frac{1 {\rm
Hz}}{f}\right) \, .
\eeq
The effective noise from all NS-NS inspirals (before subtraction) is plotted in Fig.~2,  alongside 
the noise level from the sought-for inflationary background and BBO's instrumental noise curve.
Clearly, the NS-binary foreground has amplitude $\sim 10^2$ times higher than the (hypothetical)
inflationary background's, in the BBO band, and so it must be possible to reduce (by subtraction)
the foreground amplitude by more than  $\sim 10^{2.5}$ to reveal an underlying primordial
background.

\begin{figure}[ht]
\centerline{\includegraphics[width=9cm]{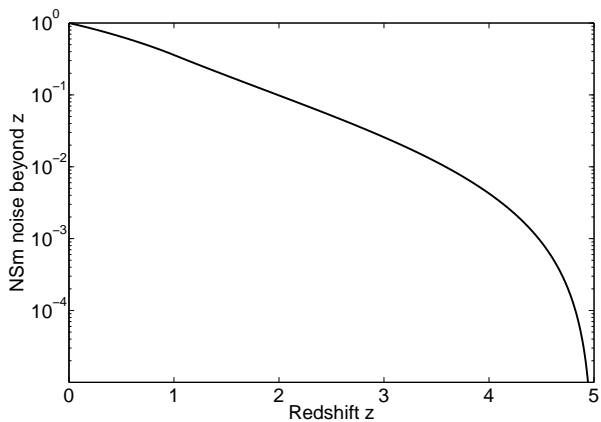}}
\caption[Fractional NS/NS foreground noise]{ Figure
plots $S_{\rm h}^{\rm NSm, >z}/S_{\rm h}^{\rm NSm}$ vs. $z$, i.e., it plots the
fractional contribution of NS-NS binaries beyond redshift $z$ to the total NS-NS foreground noise. }
\label{figFracNSNS}
\end{figure}

Given our $r(z)$ and fiducial cosmological model,
it is also straightforward
to determine what fraction of $S_{\rm h}^{\rm NSm}(f)$ is due to sources farther
out than some given redshift $z$. The result is plotted in
Fig.~\ref{figFracNSNS}. For example, $64\%$ of the foreground spectral density is due to sources at $z< 1$, and 
$99\%$ is due to sources merging at $z < 3.6$.
Thus, very roughly speaking, one must subtract out
all NS-NS mergers up to $z\approx 3.6$ to reduce the foreground noise {\it amplitude}
by one order of magnitude.
Of course, that conclusion  is too simplistic, since
the redshift out to which any particular NS binary can be observed depends
on that binary's orientation
as well as its redshift; see Section VI
below for a proper accounting of this dependence.\newline

\section{Understanding confusion noise: Why NS-NS chirps interfere with each other}

So far, we have computed a spectrum for the NS-NS inspiral foreground, but
we have not yet explained in what sense this foreground represents a noise source for BBO.
We do so in this section, showing how GW signals from different mergers
``interfere with'' and so obscure each other.
In this paper we simply sketch the main results; full details will be provided elsewhere~\cite{Cut2006}.

\subsection{Brief Review of Optimal Matched Filtering}
Typical NS-NS merger signals will have amplitudes roughly two orders of magnitude smaller than
the amplitude of BBO's instrumental noise. In practice, therefore, (some version of) matched filtering will
be required to dig these buried signals out
of the noise. Hence we will begin by briefly reviewing optimal matched filtering,
partly to fix notation. For a more complete
discussion, see  \cite{CuFl1994} or \cite{WaZu1962}.

The output of $N$ detectors can be represented by the
vector $s_{\rm A}(t)$, $A = 1,2,...,N$. It is often convenient
to work with the Fourier transform of the signal; the convention we use is
\begin{equation}
\label{fourierT}
{\tilde s}_{\rm A}(f) \equiv \int_{-\infty}^{\infty}\, e^{2\pi i f t}s_{\rm A}(t)\, \drm t,
\end{equation}

The output $s_{\rm A}(t)$ is the sum of gravitational wave signal 
$h_{\rm A}(t)$ plus instrumental noise $n_{\rm A}(t)$.
In this section we will assume that the instrumental noise is both stationary and Gaussian.
`Stationarity' essentially means that the different Fourier components
$\tilde n_{\rm A}(f)$ of the noise are uncorrelated; thus we have
\begin{equation}
\label{pr}
\overline{ {\tilde n_{\rm A}}(f) \, {\tilde n_{\rm B}}(f^\prime)^*}  = \frac{1}{2}
\delta(f - f^\prime) S_{{\rm h},{\rm AB}}(f),
\end{equation}
where an overline `$\overline{\phantom{abc}}$' denotes the `expectation value' and
$S_{{\rm h},AB}(f)$ is referred to as the spectral
density of the noise.
[When N=1 (i.e., when there is just a single
detector), we will dispense with detector indices and
just write $\tilde s(f)$ and $S_{\rm h}(f)$.]
For our problem, we can restrict
attention to the case where noises in different detectors are uncorrelated; then we have
\begin{equation}
\label{prUncorr}
\overline{ {\tilde n_{\rm A}}(f) \, {\tilde n_{\rm B}}(f^\prime)^* } = \frac{1}{2}
\delta(f - f^{\prime}) S_{{\rm h},{\rm A}}(f) \delta_{\rm AB} \, .
\end{equation}

Given stationarity, `Gaussianity' implies that each Fourier component has
Gaussian probability distribution.
Under the assumptions of stationarity and Gaussianity, we obtain a
natural inner product on the vector space of signals.
Given two signals $g_{\rm A}(t)$
and $k_{\rm A}(t)$, we define $\langle {\bf g} \, | \, {\bf k} \rangle$ by
\begin{equation}
\label{inner}
\langle {\bf g} \,|\, {\bf k} \rangle = 2 \sum_{\rm A} \int_{-\infty}^{\infty} \frac{\tilde g_{\rm A}^*(f) \tilde k_{\rm A}(f) \, \drm f}{S_{\rm h,A}(f) } \, .
\end{equation}
It also follows from Eqs.~(\ref{prUncorr}) and (\ref{inner}) that for any functions $g_{\rm A}(t)$
and $k_{\rm A}(t)$,
the expectation value of $({\bf g}|{\bf n}) ({\bf k}|{\bf n})$,
for an ensemble of realizations of the
detector noise $n_{\rm A}(t)$, is just $({\bf g}|{\bf k})$.

In terms of this inner product, the probability for the
noise to have some realization
${\bf n}_0$ is just
\begin{equation}
\label{pn0}
p({\bf n} = {\bf n}_0) \, \propto \, e^{- \langle {\bf n}_0\, |\, {\bf n}_0
\rangle /2}.
\end{equation}
Thus, if the actual incident waveform is ${\bf h}$, the probability of measuring a signal ${\bf s}$ in the
detector output is proportional to
$e^{-\langle {\bf s-h} \, | \, {\bf s-h}\rangle/2}$.  Correspondingly, given a measured signal ${\bf s}$, \
the
gravitational waveform ${\bf h}$ that ``best fits'' the data is the one that
minimizes the quantity $\langle {\bf s-h} \, | \, {\bf s-h} \rangle$.

For a given incident gravitational wave, different realizations
of the noise will give rise to somewhat different best-fit
parameters.  However, for large SNR, the best-fit parameters will have a
Gaussian distribution centered on the correct values.
Specifically, let ${\tilde \lambda}^\alpha$ be the ``true'' values of the
physical parameters,
and let ${\tilde \lambda}^\alpha + \Delta \lambda^\alpha$ be the best
fit parameters in the presence of some realization of the noise.  Then
for large $\rm SNR$, the parameter-estimation errors $\Delta \lambda^\alpha$ have
the Gaussian probability distribution
\begin{equation}
\label{gauss}
p(\Delta \lambda^\alpha)=\,{\cal N} \, {\rm e}^{-\frac{1}{2}\Gamma_{\alpha\beta}\Delta \lambda^\alpha
\Delta \lambda^\beta}.
\end{equation}
Here $\Gamma_{\alpha\beta}$ is the so-called Fisher
information matrix defined
by
\begin{equation}
\label{sig}
\Gamma_{\alpha\beta} \equiv \left\langle \frac{\partial {\bf h}}{\partial \lambda^\alpha}\, \Big| \,
\frac{\partial {\bf h}}{\partial \lambda^\beta }\right\rangle \,
\end{equation}
and ${\cal N} = \sqrt{ {\rm det}({\bf \Gamma} / 2 \pi) }$ is the
appropriate normalization factor.  For large $\rm SNR$, the
variance-covariance matrix is given by
\begin{equation}
\label{bardx}
\overline{\Delta \lambda^\alpha \Delta \lambda^\beta}
= (\Gamma^{-1})^{\alpha\beta} + {\cal O}({\rm SNR})^{-1} \;.
\end{equation}

In the above notation,
optimal filtering for some gravitational-waveform $h(t)$ simply amounts to
taking the inner product of $h(t)$ with the data stream $s(t)$.
Assuming ${\bf s} = {\bf n} + {\bf h}$, then
\be\label{sdoth}
\langle {\bf s} \, | \,  {\bf h} \rangle
= \langle {\bf n} \, | \,  {\bf h} \rangle + \langle {\bf h} \, | \,  {\bf h} \rangle
\ee

The first term on the rhs of Eq.~(\ref{sdoth}) has rms value
$\langle {\bf h} \, | \,  {\bf h} \rangle^{1/2}$, so the signal-to-noise of the detection will be
approximately given by
\begin{equation}
\label{snh}
{\rm SNR}[ {\bf h}] =
\frac{\langle{\bf h}|{\bf h}\rangle}{{\rm rms}\  \langle{\bf h}|{\bf n}\rangle} =
\langle{\bf h}|{\bf h}\rangle^{1/2}.
\end{equation}

\subsection{Overlapping NS-NS chirps as a source of self-confusion}

Now imagine that the detector output $s(t)$ consists of instrumental noise $n(t)$ plus
the sum of some large number of merger signals (labelled by ``i''):
\beq
s(t) = n(t) + \sum_i h_i(t) \, .
\eeq
(For simplicity,  here we will consider the case of a single detector, and so eliminate the index $A$; the generalization to
multiple detectors is
trivial.)

As explained above, optimally filtering the data for any particular merger waveform $h_j(t)$ is
equivalent to taking the inner product $\langle {\bf s} \, | \,  {{\bf h}_j} \rangle $, which we can
write as the sum of three pieces:
\beq\label{3terms}
\langle {\bf s}\, |\, {\bf h}_j \rangle = \langle {\bf n}\, |\, {\bf h}_j \rangle  +
\sum_{i \ne j} \langle {\bf h}_i\, |\, {\bf h}_j \rangle
+ \langle {\bf h}_j\, |\, {\bf h}_j\rangle  \, .
\eeq

For the signal to be detectable, the third term should be significantly larger than the rms values
of the first and  second terms.
We now explain {\it why} the second term can be sizeable; i.e., why different chirp signals can have
substantial overlaps~\footnote{One of the authors, C.C., presented this analysis in Dec., 2001 at  GWDAW2001 in Trento, Italy, but heretofore had not published it.}. 
To simplify this discussion, let us use a slightly simpler version of the inner product; define
\beq\label{newinner}
( {\bf g} \, | \,{\bf k} )
\equiv   \int_{-\infty}^{\infty}\tilde g^*(f) \tilde k(f) \drm f =  \int_{-\infty}^{\infty}g(t) k(t) \drm t \, ,\eeq
where the second equality in Eq.~(\ref{newinner}) is just Parseval's theorem.
(Clearly, this is just our usual inner product, but without the "re-weighting by $1/S_{\rm h}$" in the frequency domain.
For white noise, where $S_{\rm h}(f) = \rm constant$,
$( \, | \, )$ and $\langle\, | \, \rangle$ are equivalent, except for an overall constant.)

We now want to estimate the values of $( {\bf n} \, | \,{\bf h}_i  )$ and $( {\bf h}_i \, | \,{\bf h}_j )$
for any two binary inspiral waveforms $h_i(t)$ and $h_j(t)$.
In the nearly-Newtonian regime of interest to BBO, these are simple chirp waveforms:
\bea
h_i(t) &=& A_i(t) \cos\Phi_i(t)  \, , \\
h_j(t) &=& A_j(t) \cos\Phi_j(t)  \, ,
\eea
where
\bea
\Phi_i(t) &=& \cos\int^t2\pi f_i(t') \drm t'  \, , \\
\Phi_j(t) &=& \cos\int^t2\pi f_j(t') \drm t'  \, ,
\eea
and where $A_i(t)$, $A_j(t)$, $f_i(t)$ and $f_j(t)$ are
all slowly varying (meaning their fractional change during one cycle is $<<1$), and
$f_i(t)$ and $f_j(t)$ are  monotonically increasing.
Then, since the integrand is highly oscillatory, it is clear that
the integral $\int{h_i(t) h_j(t) \, \drm t} $  will show substantial waveform overlap only
if there is some instant $t_0$ when the two signals have the same frequency:
\beq
f_i(t_0) = f_j(t_0)  \, .
\eeq
I.e., if one considers the "track" of each signal in the f-t plane, then $t_0$ is the
instant of time when the two tracks cross. Using the stationary phase approximation,
it is straightforward to show that~\cite{Cut2006}:
\be\label{overlap}
(\, {\bf h} _i \,|\, {\bf h}_j ) \approx \frac{1}{2} A_i(t_0) A_j(t_0) |\delta \dot f|^{-1/2}  \cos[\Delta\Phi_0 \pm \pi/4] \, ,
\ee
where $\Delta \Phi_0 \equiv [\Phi_i(t_0) - \Phi_j(t_0)]$,
 $\delta \dot f
\equiv  [\dot f_i(t_0) - \dot f_j(t_0)]  $, and where the sign in front of
the $\pi/4$ in Eq.~(\ref{overlap}) is positive when $\delta \dot f > 0$ and negative when
$\delta \dot f < 0$.

We want to use this result to estimate
\be
\Big({\bf h}_j \, \Big| \, \sum_{i \ne j}  {\bf h}_i \, \Big) \, ,
\ee
ie., to sum the contributions from all binaries whose f-t tracks overlap the $j^{\rm th}$ track.
Since the phase differences $\Delta \Phi_0$
at different intersections will clearly be uncorrelated, the contributions accumulate
in a random-walk fashion; i.e., the square of the sum  is approximately the sum of the squares of the
individual terms.
Also, as we show in the next subsection, a typical NS-NS ``track'' will intersect
a very large number of tracks from other merging binaries, so we are in the realm
of large-number statistics.
Finally, while the magnitude of each squared-contribution scales like
$|\delta \dot f|^{-1}$, the number of terms in the sum scales like the average value of
$|\delta \dot f|$, since the larger the ``relative velocities'' of the tracks, the more crossings.
The dependence of the sum on the typical size of $|\delta \dot f|$ therefore ends up cancelling out, and
one can show the following~\cite{Cut2006}.
Let $H(t) = \sum_i h_i(t)$ be the entire foreground
generated by NS-NS chirps,  and let $H$'s spectral density be $S_{\rm H}(f)$, normalized so that
\be
\overline{H^2(t)} = \int_0^\infty{S_{\rm H}(f)\, \drm f} \, .
\ee
Then the expectation value of $({\bf h}_j \, | \, \sum_{i \ne j}  {\bf h}_i \, )^2 $ is given by
\be
\overline{\Big({\bf h}_j \, \Big| \, \sum_{i \ne j}  {\bf h}_i \, \Big)^2} = \frac{1}{2} \int h^2_j(t) S_{\rm H}\big(f_j(t)\big)\, \drm t \, .
\ee
But the same result holds for the mean-square overlap of $h_j(t)$ with stationary, Gaussian noise
$n(t)$:
\be
\overline{({\bf h}_j \, | \, {\bf n} \, )^2} = \frac{1}{2} \int h^2_j(t) S_{\rm h}\big(f_j(t)\big)\, \drm t
\ee
with $\overline{n^2(t)} = \int_0^\infty{S_{\rm h}(f)\, \drm f}$. I.e., the mean-square overlap of a single chirp $h_j(t)$
with the chirp foreground
$H(t)$ (excluding  $h_j$ itself) is the same as the mean-square overlap of $h_j(t)$ with stationary, Gaussian noise
having the same spectral density as $H$.
(It is straightforward to generalize this result to inner products with
non-trivial frequency-weighting~\cite{Cut2006}.)
It is for this reason that in Eq.~(\ref{eq:tot1}) we simply add together the spectral densities of the
instrumental noise and the ``confusion noise'' from unresolved chirps.

\subsection{The number of overlapping inspiral tracks in the f-t plane}

We saw in the previous subsection that two chirp signals have substantial overlap
only if their tracks in the f-t plane intersect.
Here we consider the track from a typical NS-NS inspiral and estimate
how many {\it other} inspiral tracks it crosses.

Let $\rho(f)$ be the probability density of merger signals in frequency space;
i.e.,  at any instant, $\rho(f) \Delta f$ is the average number of NS-NS GW signals  received near
the Earth that are in the frequency range
$[f - \Delta f/2, f + \Delta f/2]$.  Since the BBO mission lifetime is vastly shorter
than the age of the universe, we can assume $\rho(f)$ is time-independent, implying
\beq
\rho(f) \frac{\drm f}{\drm t} = {\rm const} = \dot N \, ,
\eeq
where, again, $\dot N \equiv \Delta N_{\rm m}/\Delta \tau_0$ is the total rate of mergers
in the observable universe (from all $z$).
The GW frequency derivative $\dot f$ is given by
\beq\label{dfdt}
\drm f/\drm t = \frac{96}{5}\pi^{8/3}\,\left[{\cal M}(1+z)\right] ^{5/3} \, f^{11/3}\, .
\eeq
so clearly $\rho(f) \propto f^{-11/3}$.

Now consider any one track in the $f-t$ plane, and examine it in the neighborhood of some frequency $f$.
It is easy to see that the rms rate $r_{\rm c}$ at which it intersects neighboring tracks is
\beq\label{cross}
r_{\rm c} = 0.5 \,\rho(f) \, \Delta \dot f
\eeq
where
$\Delta \dot f$ is the rms variation in frequency derivatives for
sources with GW frequency $f$.
The $0.5$ factor in Eq.(\ref{cross}) arises because, for any two neighboring tracks at any instant, there
is a $50\%$ chance that they are approaching each other and a $50\%$ chance that they are separating.

Using Eq.~(\ref{dfdt}),
we see that the rms relative ``velocity'' of nearby tracks is
\beq
\frac{\Delta \dot f}{\dot f}  = 5/3 \frac{\Delta {\cal M}_{\rm eff}}{{\cal M}_{\rm eff}} \, ,
\eeq
where we define ${\cal M}_{\rm eff} \equiv {\cal M}(1+z)$, and
where $\Delta {\cal M}_{\rm eff}$ is the rms variation in this quantity.
Now the fractional variation in $\cal M$ itself, $\Delta {\cal M}/{\cal M}$, is probably
at least of order $0.1$. However, from Fig.~\ref{figNSNSnum} we see that this is small compared to the
variation $\Delta(1 +z)/(1+z)$, which is $\sim 0.4$.

Thus $r_{\rm c}$ is roughly given by
\beq
r_{\rm c}  = 0.5 (5/3) \rho \dot f \frac{\Delta {\cal M}_{\rm eff}}{ {\cal M}_{\rm eff}} \sim \frac{1}{3} \dot N \, 
\eeq
independent of the particular frequency $f$.  That is, the rate at which any particular track crosses all other
tracks is about one-third the total merger rate from all observable sources,  independent of where
one is on the track. Thus, for any one track over the last 3 years of inspiral, one expects of order $10^5$ crossings.
This amply justifies our use of large-number statistics in the previous subsection.

\section{Confusion noise from imperfectly subtracted waveforms}

NS binaries limit BBO's sensitivity to a primordial background in two ways.
First, there will be some binaries that are too weak (because of their
distance and/or orientation) to be individually identified and subtracted, and
these ``unidentified binaries'' clearly represent a source of ``confusion noise.''
Second, even identified NS binaries will not be removed perfectly from the data stream;
inevitably (due to  the finite signal-to-noise of the observations) there
are subtraction errors, which represent a second source of confusion noise.
This section addresses the confusion noise that results from subtraction errors.
First we will prove a simple theorem regarding the magnitude of subtraction errors.
Then we will sketch a simple strategy for largely eliminating their impact on
other analyses by projecting them out, at the cost of some bandwidth.
We estimate that lost bandwidth for BBO, and conclude that the
loss is small enough that in the rest of this paper we can safely neglect it.

We believe the analysis and strategy we outline here will also be useful in similar
contexts, especially in dealing with problems of confusion noise in LISA data.
Here we provide only a sketch of the main ideas; more details will be provided in a forthcoming
publication~\cite{Cut2006}.

\subsection{Subtraction errors due to noise}
We have argued that, before searching for a primordial GW background, one
will want to first subtract from the data the best fit to each identified inspiralling
compact binary. However, because of detector noise, the best-fit values of the
binary parameters will differ from their true values, and so
the best-fit waveforms will be somewhat in error. What is the typical size of the error?
That is easy to calculate:
Let $h(t)$ be some gravitational waveform immersed in noisy data, and assume the waveform depends on
$N_{\rm p}$ physical parameters $\lambda^{\alpha}$ ($\alpha = 1,\cdots,N_{\rm p}$).
Because of the noise, the best-fit parameter values
$\hat \lambda^{\alpha}$ will differ from the true parameter values by \cite{CuFl1994}
\be
\delta \lambda^{\alpha} \equiv \hat \lambda^{\alpha} - \lambda^{\alpha}  \approx \big(\Gamma^{-1}\big)^{\alpha \beta}\langle {\bf n} \, | \, \partial_{\beta} {\bf h}\rangle \, ,
\label{eqParaErr}
\ee
and, correspondingly,
the best-fit waveform $\hat h(t)$  will differ from the true one by
\beq
\begin{split}
{\bf \delta h} &\equiv \bf{\hat h - h}  \\
&= \partial_{\alpha} {\bf h} \, \delta \lambda^{\alpha} + {\cal O}(\delta \lambda)^2 \, .
\end{split}
\label{eqDiffBF}
\eeq
Using Eqs.(\ref{sig}), (\ref{bardx}), and ~(\ref{eqDiffBF}), we can immediately estimate the norm-squared of this residual error. To lowest
order in $\delta \lambda^{\alpha}$, we have
\beq
\begin{split}
\overline{\langle {\bf \delta h} \, | \, {\bf \delta h}\rangle }  &= \langle \partial_{\alpha} {\bf h} \, | \, \partial_{\beta} {\bf h} \rangle \,
\overline{\delta \lambda^{\alpha} \delta \lambda^{\beta}} \\
&= \Gamma_{\alpha\beta}\big(\Gamma^{-1}\big)^{\alpha\beta} = N_{\rm p} \, .
\end{split}
\label{eqNSofRes}
\eeq
Thus the size of $\langle \delta {\bf h} \, \big| \,  \, \delta {\bf h} \rangle$
is independent of the signal strength, but increases
linearly with the number of parameters that need to be fit for.

Eq.~(\ref{eqNSofRes}) estimates the weighted integral of $|\delta \tilde h(f)|^2$; it says nothing about
rms size of $|\delta \tilde h(f)|^2$ at any {\it particular} frequency $f$.
Now, one can always calculate $|\delta \tilde h(f)|^2$ using (to lowest order)
\beq
|\delta \tilde h(f)|^2 = \partial_{\alpha} \tilde h(f) \partial_{\beta} \tilde h^*(f) \, (\Gamma^{-1})^{\alpha\beta}\, ,
\eeq
but for back-of-the-envelope calculations, it is reasonable to simply turn Eq.~(\ref{eqNSofRes}) into a point estimate for the
relative error:
\be
\frac{|\delta \tilde h(f)|}{|\tilde h(f)|} \sim \left[
\frac{\langle \delta {\bf h} \, \big| \,  \delta {\bf h} \rangle}{\left< {\bf h} \,
\big| \,  {\bf h} \right>} \right]^{1/2} \sim \frac{N_{\rm p}^{1/2}}{\rm SNR} \, .
\ee
For BBO measurements of NS-NS binaries, $N_{\rm p} \approx 11$ (cf. Section VI), and 
for a typical source (i.e, for a source at $z \approx 1.5$, with $\mu = 0.5$, where
$\mu$ is the cosine of the angle between the line-of-sight and the normal to the binary's
orbital plane), $\rm SNR \approx 140$, 
so $\delta h/h \sim 2.4 \times 10^{-2}$.

Given the extreme accuracy with which foreground sources must be subtracted, at first glance this level of error
seems unacceptable. However
it would be a mistake to regard  $\delta h$ as a completely random, additive noise source in the data.
For one thing, after the best-fit signal $\hat h(t)$ has been removed from the data stream,
the amplitude of noise plus residual is {\it smaller} (on average) than that
of the noise alone.
To see this, consider again the case of data 
$s(t) \equiv  n(t) + h(t)$, and assume that the observation time is $T$, and that the data has been band-limited
to $[- f_{\rm max}, f_{\rm max}]$. Then it is easy to show that the noise alone has squared-magnitude:
\be
\overline{\langle {\bf n} \, | \, {\bf n} \rangle} = 2 f_{\rm max} T \, ,
\ee
which is just the number of data points, for data sampled at the Nyquist rate $2 f_{\rm max}$.
Next consider the magnitude-squared of the post-subtraction data set, ${\bf s} - \hat{\bf h}$, where again $\hat{\bf h}$ is the waveform
that best fits the data. 
A straightforward calculation shows that~\cite{Cut2006}
\bea
\overline{\langle {\bf s} - \hat{\bf h} \, \big| \,  \, {\bf s} - \hat{\bf h}\rangle} &=&
\overline{\langle {\bf n } - \delta {\bf h}
\, \big| \,  {\bf n }   - \delta {\bf h} \rangle}  \\
&=& 2 f_{\rm max} T \,   - N_{\rm p} \, .
\eea
I.e., fitting out waveform $\hat h$ causes the norm-squared of the data to decrease below
what is expected from noise alone.
This  is easy to understand: the fitting procedure takes out not only the signal ${\bf h}$, but
also that part of the noise that "looks like" the difference between ${\bf h}$ and some other
waveform, $\hat{\bf h}$, having slightly different physical parameters.  Stated geometrically, if one
considers the manifold of physical gravitational waveforms, embedded in the
vector space of possible measured signals, one sees that any piece of the noise that
is tangent to the waveform manifold (at the location of the true signal) gets fitted out.
Indeed, one sees from Eq.~(\ref{eqParaErr}) that it is just this piece of the noise, lying in the tangent space
to the waveform manifold, that is ``responsible'' for the
parameter estimation errors $\delta \lambda^{\alpha}$ in the first place.
In the next subsection we outline a strategy projecting out this error
{\it before} one searches for an inflation-generated background.

Note that nothing in the above arguments required the signal to emanate from a single physical source.
E.g., if $H(t)$ is the entire foreground signal coming from $N_{\rm s}$ sources,
\be
H(t) = \sum_{i=1}^{N_{\rm s}} h_i(t) \, ,
\ee
and if each $h_i(t)$ is described by $p$ parameters, then
the full parameter space is described by $N_{\rm p} =  p \times N_{\rm s}$ parameters, and
\be\label{dH}
\overline{\langle {\bf \delta H} \, | \, {\bf \delta H}\rangle } =  p \times N_{\rm s} \,
\ee
to lowest order in $1/\rm SNR$.  The total $\rm SNR^2$ of the foreground $H$ is just
$N_{\rm s}$ times the average $\rm SNR^2$ of the individual sources, and $N_{\rm p}$ is of course
directly proportional to $N_{\rm s}$, so the fractional error in subtracting the whole
foreground is just the fractional error in subtracting a typical source:
\be
\delta H/H \sim \delta h/h \, .
\ee
For BBO measurements of the NS-NS foreground, we thus estimate
$\delta H/H \sim 2.4 \times 10^{-2}.$

As a digression, we remark that because
our foreground consists of a large number of overlapping sources, it should not be surprising if there
are some near-degeneracies that make it practically impossible to determine some of the physical parameters
of some of the sources.  (These are cases where the affect on $H(t)$ of adjusting the parameters of one source can be almost
perfectly cancelled by adjusting the parameters of another source.)
We bring this up to make the point that such near-degeneracies do {\it not} necessarily
imply any degradation in one's ability to subtract out the foreground. Indeed, in the case of very high SNR (per source),
it implies the opposite: near degeneracies would imply that the residual ${\bf \delta H}$  is somewhat smaller than
estimated above.
The reason is simple: a near degeneracy means
that the effective dimensionality of the
signal space (near the actual signal) is smaller than the number of parameters being used to describe it.
I.e., one could find a new parametrization using a fewer number of variables, $N^{\prime}_{\rm p}$.
Then a repetition of the above arguments would yield
$\overline{\langle {\bf \delta H} \, | \, {\bf \delta H}\rangle } =  N^{\prime}_{\rm p} < N_{\rm p}$.
For the BBO case, where SNR per source is $\approx 140$, it probably will require detailed simulations
to determine whether subtraction errors are larger or smaller than indicated by the high-SNR result,
Eq.~(\ref{dH}).  We leave this question to future work.

\subsection{Projecting out residual subtraction errors}
In this subsection we propose one strategy for effectively cleaning
the BBO data of subtraction errors, after the NS-NS binaries have been subtracted out.
Using this strategy, we argue that the impact of subtraction residuals (arising from instrumental noise)
becomes sufficiently small that they can be ignored in the rest of this paper.
We do not argue that our strategy is the best one possible, but rather offer
it as an ``existence argument'' that some such strategy is possible.
The use of any alternative strategy that leads to the same conclusion
would not affect the main results of this paper.

The basic observation behind our strategy is that the residual $\delta H(t)$ is
mostly confined to a surface within the vector space of all signals: the tangent
space to the waveform manifold at the best-fit point.
The corresponding errors in the subtracted waveform can be expanded in a Taylor series:

\beq
\delta H(t) = \partial_{\alpha}H(t) \delta\lambda^{\alpha}  +
\frac{1}{2} \partial_{\alpha}\partial_{\beta}
H(t) \delta\lambda^{\alpha} \delta\lambda^{\beta}  + \cdots
\eeq
where $\alpha, \beta = 1,...,p\times N_{\rm s}$.
The first-order piece on the rhs is the linear combination of $N_{\rm p} = p\times N_{\rm s}$ wavefunctions
(the $\partial_{\alpha}H(t)$),  with unknown coefficients (determined by the noise).
We propose projecting these directions out of the data stream.
This is simple in principle. Consider the operator
\be
P \equiv I - (\Gamma^{-1})^{\alpha\beta}|\partial_{\alpha}{\bf H}\rangle \langle \partial_{\beta}{\bf H}\big| \, .
\ee
where for simplicity we use here standard bra-ket notation of quantum mechanics.
It is trivial to verify that $P^2 = P$ and that $P$ destroys any wavefunction of the
form $\partial_{\alpha}H(t) \delta\lambda^{\alpha}$.
We propose acting on the data streams with $P$ before searching them for an inflation-generated background.

What fraction of the data have we thrown away, by using $P$?
For a fiducial 3-yr BBO lifetime, with, say,  $\sim 3\times 10^5$ subtracted sources, each
determined by $\sim 11$ parameters, $N_{\rm p} \sim 3 \times 10^6$. Assuming a 2-Hz sampling rate 
(sufficient for capturing most of the signal),
with $\sim 10^8\,$s of data and $8$ independent channels, the dimension S of the full data
space is $S \sim 1.5 \times 10^9$. Thus the fraction of the data that is discarded is only
$N_{\rm p}/S \sim 2 \times 10^{-3}$, which is a negligible loss.

So far we have discussed projecting out the first-order piece of the
subtraction error; i.e., the piece linear in the parameter estimation errors
$\delta \lambda^\alpha$.  What is the magnitude of the second-order subtraction errors (i.e., the
ones quadratic in $\delta \lambda^\alpha$)?  This is clearly given by
\be\label{d2H}
\overline{\langle {\delta^2 {\bf H}} \, | \, {\delta^2 {\bf H}}\rangle }  = 
\frac{1}{4}\langle \partial_{\alpha} \partial_{\beta} 
{\bf H} \, | \, \partial_{\gamma} \partial_{\epsilon} {\bf H} \rangle \,
\overline{\delta \lambda^{\alpha} \delta \lambda^{\beta} \lambda^{\gamma} \delta \lambda^{\epsilon}}  \, .
\ee
but evaluating the rhs of Eq.~(\ref{d2H}) is beyond the scope of this paper, and so we
content ourselves with a cruder estimate. The second-order errors clearly scale like the
square of the first-order errors, so 
a very crude estimate is $\delta^2 H/H \sim (\delta H/H)^2 \sim 6 \times 10^{-4}$.  Of course, 
this estimate is properly multiplied 
by some pre-factor (which can only be obtained by calculating of the rhs of 
Eq.~\ref{d2H} ). 
Depending on this pre-factor and the actual level of the NS-NS foreground, these 
second-order subtraction errors could be comparable in size to the sought-for inflationary 
background. If this is the case, we would advocate projecting out the second-order errors as well.
The second-order errors are linear combinations of second derivatives $\partial_{\alpha}\partial_{\beta} H(t)$.
It is important to notice that such second derivatives vanish identically
{\it unless $\alpha$ and $\beta$ are parameters describing the same binary}.
Thus the vast majority of such second derivatives vanish. 
For each binary, there are $(11\times 12)/2 = 66$ non-vanishing second derivatives, so projecting out
the second-order piece of the subtraction errors would cost only $\sim 1 \%$ of BBO's bandwidth.
A crude estimate of the size of third-order  subtraction errors is
$\delta^3H/H \sim (\delta H/H)^3 \sim 10^{-5}$.  Clearly, unless the missing pre-factor here
is quite large (of order $100$ or more), it should not be necessary to project these third-order errors 
out of the data. 

\section{Catalog of relevant physical parameters and relevant effects}

\subsection{Subtraction errors due to inaccurate waveform templates}

In the previous section, we outlined a method for handling subtraction errors arising from
instrumental noise. Another potential source of subtraction error is inaccurate theoretical
template waveforms. Provisionally, we will  
regard a physical parameter, effect, or post-Newtonian term 
as ``relevant for BBO'' if neglecting it would lead to relative errors in our
theoretical inspiral waveforms of size $\delta h/h \gtrsim 10^{-3}$ (since
errors of that magnitude could dominate over the inflationary background).
Since each inspiral waveform contains $\sim 10^7$ cycles, 
knowing  the waveforms to 
$\delta h/h \gtrsim 10^{-3}$ requires calculating the waveform phase to 
roughly one part in $10^{11}$!

The post-Newtonian (PN) expansion is clearly the right tool for constructing the
waveforms, since the PN
expansion parameter $M/r$ is small in the BBO band:  
\beq\label{Mr}
\frac{M}{r} \approx 5.5 \times 10^{-4} \left(\frac{M[1+z]}{2.8 M_{\odot}}\right)^{2/3}\left(\frac{f}{0.3\,{\rm Hz}}\right)^{2/3} \, ,
\eeq
where $f$ is the GW frequency.
If one uses PN waveforms, the only reasons for theoretical error would be
1) failure to calculate post-Newtonian corrections to sufficiently high order in the PN 
expansion, or 2) failure to account for all relevant physical parameters (e.g., the
spins of the NSs).

This section provides an initial ``scoping out'' of the questions of which physical parameters
are relevant, and which post-Newtonian order is sufficient.  

\subsection{Orbital Eccentricity}
\subsubsection{Typical eccentricities of binaries in the BBO band}

Here we consider the implications of small (but non-zero) eccentricity for
the subtraction problem. We begin by estimating typical eccentricities of 
NS binaries when they are emitting GWs in the BBO band.

It is well known that radiation reaction tends to circularize the
orbits of nearly Newtonian binaries. For small eccentricity $e$,
$e^2$ decreases with the orbital period $P$ according to 
$e^2 \propto P^{19/9}$~\cite{Pet1964}.  For arbitrary $e$, the mutual 
scaling is given by~\cite{Pet1964}:
\beq\label{e2P}
P^{2/3} \propto \frac{e^{12/19}}{(1-e^2)}\bigg[1 + \frac{121}{304} e^2\bigg] \, .
\eeq
The two known NS-NS binaries that dominate current merger rate estimates are
PSR 1913+16 and PSR J0737-3039. 
Extrapolating from today's values of $e$ and $P$ for these two binaries, using Eq.~(\ref{e2P}), we
estimate that their eccentricities when they pass through the BBO band
will be
$e^2_{1913} \approx 4.6 \times 10^{-8} \left(\frac{f}{0.3\,{\rm Hz}}\right)^{-19/9}$ and $ e^2_{0737} \approx 2.0 \times 10^{-9} \left(\frac{f}{0.3\,{\rm Hz}}\right)^{-19/9}$.
Based on these two examples,  we will provisionally assume that
typical eccentricities are 
$e^2 \sim [10^{-9}-10^{-7}]\left(\frac{f}{0.3\,{\rm Hz}}\right)^{-19/9}$.
However we will also consider the implications of a subpopulation of 
NS binaries with considerably larger eccentricity.

\subsubsection{Effect of non-zero eccentricity on waveform phase}
The effect of small, non-zero eccentricity is to
slightly increase the inspiral rate; to lowest nontrivial PN order and to first
order in $e^2$, the increase 
(derivable from \cite{Pet1964}) is given by:
\be
\drm f/\drm t = \drm f/\drm t|_{e=0}\left[1 + \frac{157}{24} e^2\right]
\ee

In the stationary phase approximation, we can write the Fourier transform
of the emitted waveform (omitting tensor indices) as~\cite{CuFl1994}
\be\label{statphase}
\tilde h(f) \propto
\big({\cal M}(1+z)\big)^{5/6}f^{-7/6}[1 + \ldots] \e^{\irm \Psi(f)} \, ,
\ee
where ``$\ldots$'' stands for higher-order
PN corrections, and where the phase $\Psi(f)$ can be written as
\beq
\Psi(f) = \Psi_0(f) + \Psi_{\rm e}(f) \, .
\eeq
Here $\Psi_0(f)$ represents the zero-eccentricity phase evolution and
has the following PN expansion:
\beq
\label{pnpsi}
\begin{split}
\Psi_0(f) = &
\;{\rm const} + 2\pi f t_{\rm c} +\frac{3}{4}\left(8\pi {\cal M}(1+z) f \right)^{-5/3} \\
\mbox{} &  \times \, \left[1 + \frac{20}{9}\left(\frac{743}{336}+\frac{11\mu}{4M}\right) y -16 \pi y^{3/2} + \ldots \right].
\end{split}
\eeq
with $y \equiv (\pi M (1+z)\, f )^{2/3}$, while
$\Psi_{\rm e}(f) $  represents the phase correction due to non-zero $e^2$, and
is given (again, to lowest nontrivial PN order and to first order in $e^2$) 
by~\cite{KKS1995}:
\beq\label{Psie}
\Psi_{e}(f) = -\frac{7065}{187136}\left[\pi {\cal M} (1+z)\right]^{-5/3} e_0^2 f_0^{19/9} f^{-34/9} \, .
\eeq
Here $e_0$ is the binary's eccentricity at the moment that the GW frequency 
(more specifically, the frequency of the dominant, $n=2$ harmonic)
sweeps through some fiducial frequency $f_0$. (Note that, by Eq.~(\ref{e2P}), the combination
$e_0^2 f_0^{19/9} $ is a constant, to lowest nontrivial order.)

Plugging in fiducial values,  we can re-express Eq.~(\ref{Psie}) as
\beq
\begin{split}
\Psi_{\rm e}(f) = &-0.21\\
&\times\left[\frac{e^2_{0.3\,\rm Hz}}{10^{-8}}\right] \left[\frac{(1+z){\cal M}}{1.22 M_{\odot}}\right]^{-5/3}
\left[\frac{f}{0.3\,{\rm Hz}}\right]^{-34/9} \, .
\end{split}
\label{psief}
\eeq

Note the very steep fall-off of $\Psi_{\rm e}(f)$ with increasing $f$. 
This $f^{-34/9}$ fall-off is much
steeper than for the other PN correction terms in Eq.~(\ref{pnpsi}), 
so it seems quite unlikely that errors in fitting for
$e_0$ could be ``absorbed'' into compensating errors in the other parameters.
While $\Psi_{\rm e}(f)$ is negligible for frequencies above a few Hz,
it is typically of size $\sim 2\pi$ at $f = 0.1\,$Hz.
Clearly, then, orbital eccentricity is a relevant parameter that must be accounted for,  
both in subtracting out individual sources and
in projecting out residual errors.  From Eq.~(\ref{psief}), we can also estimate roughly how accurately BBO can measure the
eccentricity of each binary; it should be possible to determine $e^2_{0.3\,\rm Hz}$ to within 
$\Delta(e^2_{0.3\,\rm Hz}) \sim [10^{-8}/{\rm SNR}] \sim 10^{-10}$.

\subsubsection{Contribution of $n=3$ radiation to $\Omega_{\rm GW}^{\rm NSm}$}

Non-zero orbital eccentricity implies that even the quadrupole piece
of the gravitational radiation is no longer purely sinusoidal, but exhibits harmonics at all
multiples $n \nu$ of the
orbital frequency $\nu$ (for integers $n \ge 1$).
Let $\dot E_{\rm n}$ be the gravitational
luminosity due to the $n^{\rm th}$ harmonic. For small $e$,
$\dot E_{\rm n} \propto e^{|2n-4|}$, so in the
range of interest for $e$, only $\dot E_3$ and $\dot E_1$ could potentially be significant.
While both $\dot E_3$ and $\dot E_1$ are $\propto e^2$, it is easy to show that the 
$n=3$ contribution to $\Omega_{\rm GW}^{\rm NSm}$ dominates over the $n=1$ contribution. Therefore
we concentrate here on the $n=3$ harmonic.

The ratio $\dot E_3/\dot E_2 $ is \cite{PeMa1963}
\be
\dot E_3/\dot E_2 \approx (3/2)^6 e^2 \, ,
\label{eqLumEcc}
\ee
from which one easily derives  
\beq
\begin{split}
\Omega&_{\rm GW}^{n\ge 3}(f)
= \bigg(\frac{3}{2}\bigg)^6 \langle e^2_{2f/3} \rangle \, \Omega_{\rm GW}^{n=2}(2f/3) \\
&= \bigg(\frac{3}{2}\bigg)^6 \bigg(\frac{3}{2}\bigg)^{13/9} \langle e^2_{f} \rangle \, \Omega_{\rm GW}^{n=2}(f) \\
& \approx 1.6 \cdot 10^{-19} \left(\frac{n_0}{10^{3}\,{\rm Mpc}^3}\right) \left(\frac{ \langle e^2_{0.3\,\rm Hz}\rangle}{10^{-8}}\right) 
\left(\frac{f}{0.3\rm Hz}\right)^{-13/9} 
\end{split}
\eeq
where  $\langle e^2_{0.3\,\rm Hz}\rangle$ is the average value (for all NS-NS mergers) of $e^2$ at $f=0.3\,{\rm Hz}$.
For our fiducial estimate of $\langle e^2_{0.3\,\rm Hz}\rangle$, this 
is significantly below the sought-for level of inflation-generated GWs, and so the extra harmonics
generated by non-zero $e$ can be neglected.

However, our estimate that $\langle e^2_{0.3\,\rm Hz}\rangle \sim 10^{-8}$ was based on the few known examples
of close NS-NS binaries; what if there is a subpopulation of NS-NS binaries that merge with substantially larger
eccentricity (e.g., due to the Kozai mechanism~\cite{Wen2003})?  The ratio of the $n=3$ to the $n=2$ piece of the waveform, 
$h^{n=3}/h^{n=2}$, is clearly of order $e$. Thus the $n=3$ piece must be subtracted (or projected out)
if $e \gtrsim 10^{-3}$.  
Fortunately, as the previous subsection makes clear, if 
$e^{\phantom{2}}_{0.3\,\rm Hz} \gtrsim 10^{-5}$, then the waveform itself will inform us of this fact, via the phase evolution of the
$n=2$ piece. 

Unfortunately, to subtract  $h^{n=3}$, one needs to know {\it both} $e$ and the  
perihelion angle $\omega$ (at some fiducial instant or frequency), since the latter clearly determines the relative phase of the 
$n=3$ and $n=2$ pieces. 
How accurately can $\omega_{0.3\, \rm\,Hz}$ be extracted from the data?  Since $\omega$ is encoded only in the
$n \ne 2$ harmonics, we estimate that $\Delta \omega_{0.3 {\rm\,Hz}} \sim {\rm min}\{\pi, (e^{\phantom{2}}_{0.3\,\rm Hz} \times {\rm SNR})^{-1}\}$.
Hence, while the $h^{n=3}$ piece is relevant for $e^{\phantom{2}}_{0.3\,\rm Hz} \gtrsim 10^{-3}$, it will be impossible to subtract it when 
$e^{\phantom{2}}_{0.3\,\rm Hz} \lesssim 10^{-2}$ (since $\omega_{0.3\,\rm Hz} $ will be undetermined).
Fortunately, even in this case,  $h^{n=3}$ can simply be projected out of the data (in the 
manner described in Sec.~IV.B) since all possible realizations of  $h^{n=3}(t)$ lie in a two-dimensional vector space.
To see this, note that if all parameters {\it except} $\omega_{0.3\,\rm Hz} $ were known, then one
could express $h^{n=3}(t)$ 
in the form $A_3(t){\rm cos}[3(\Phi_3(t) + \omega_{0.3\,\rm Hz})]$, where
$A_3(t)$ and $\Phi_3(t)$ are both known functions, and this can be expanded as
${\rm cos}[3\omega_{0.3\,\rm Hz}] \times A_3(t){\rm cos}[3\Phi_3(t)]
-{\rm sin}[3\omega_{0.3\,\rm Hz}] \times A_3(t){\rm sin}[3\Phi_3(t)]$. I.e., 
$h^{n=3}(t)$ is just some linear combination of two known waveforms, with (unknown) coefficients 
${\rm cos}[3\omega_{0.3\,\rm Hz}]$ and ${\rm sin}[3\omega_{0.3 \rm\,Hz}]$.

\subsubsection{Summary of effects of orbital eccentricity}

Extrapolating from the known NS-NS binaries, we have estimated that typical eccentricities for  
NS-NS binaries radiating in the BBO band will be $e \lesssim 10^{-4}$.
At this level, they would have a significant impact on the phase evolution of the $n=2$ harmonic, 
but the $n=3$ and $n=1$ pieces of the waveform would be negligibly small. In this case, when projecting out
residual errors, one need not worry about the perihelion angle $\omega$. 
On the other hand, if some subpopulation of NS-NS binaries has $e^{\phantom{2}}_{0.3 \rm\, Hz} \gtrsim 10^{-3}$, 
then this will be completely clear from the data itself.
For these binaries, both $e^{\phantom{2}}_{0.3\, \rm Hz}$ and  $\omega_{0.3\, \rm Hz}$ are relevant parameters, 
to be used boh in subtraction and in projecting out residual errors. Finally, there are cases when  
$\omega_{0.3\, \rm Hz}$ is relevant but impossible to determine. Fortunately, even in this case, 
$h^{n=3}$ can simply be projected out, at very modest additional cost in bandwidth. 

\subsection{Spin Effects}

We turn now to the effects of the NS spins.
Currently there are five known NS-NS binaries in our galaxy that
will merge in a Hubble time
(four binaries in the disk and one in globular cluster
M15). In only one system--PSR J0737--are the spin periods of both
NSs known. For PSR J0737, $P_{\rm A} = 22.7\,$ms and  $P_{\rm B} = 2.77\,$s.
In the other four systems, the radio-emitting neutron star is also a fast
rotator, with $P$ ranging from $28.5\,$ms to $59.3\,$ms.
The fast rotators all have low spindown rates and so appear to be recycled pulsars.
>From evolutionary considerations, one expects exactly one of
the companions to be rapidly rotating (consistent with what we find for PSR J0737).
We estimate the effect of the bodies' spins on the gravitational
waveform, for this presumed-typical case where one NS is rotating
relatively rapidly ($P \sim 30\,$ms), while the other is slowly
rotating ($P \gtrsim 1\,$s). 

\subsubsection{Precession of Orbital Plane}
If the NSs are spinning, then the orbital angular momentum vector $\vec L$ does not
have fixed direction, but instead precesses around the binary's total
angular momentum vector $\vec J$, due
to an effective $\vec L \times \vec S$ coupling.
When either 1) the two masses are nearly equal,
or 2) the spin of one NS is much greater than the other, then the
lowest-order precessional dynamics take an especially simple form--
so-called ``simple precession''~\cite{ACST1994}.
In fact, we expect both these conditions to be satisfied in most
NS-NS binaries, since (as mentioned above), we expect only one
to be rapidly rotating, and since in those binaries
where both NS masses are accurately known, the masses are indeed nearly
equal.
Therefore we shall use the simple-precession approximation to estimate
the magnitude of precessional effects on the waveform. 

Following Apostolatos et al.~\cite{ACST1994}, let $\lambda_{\rm L}$ be
the precession amplitude; i.e., the angle between $\vec J$ and $\vec L$.
While $\lambda_{\rm L}$ depends on the magnitude and direction of the
spins, the precession period depends on neither (to a very good approximation). 
The total number of precessions,
from the moment the GW frequency sweeps through $f$ until merger, is (for $M_1 \approx M_2$): 
\be\label{nprec}
N_{\rm prec} \approx 2.3 \times 10^3 \left(\frac{2.8 M_{\odot}}{M(1+z)}\right)\left(\frac{0.3 {\rm Hz}}{f}\right) \, .
\ee

It is useful to define dimensionless spin parameters $\chi_i$ by 
$\chi_i \equiv |\vec S_i|/M_i^2$.  The $\chi_i$ are related to the 
spin periods $P_i$ by
\be
\chi_i = 0.036
\left(\frac{I_i}{10^{45}\,{\rm g\,cm}^2}\right) \left(\frac{1.4 M_{\odot}}{M_i}\right)
\left(\frac{10 \, {\rm msec}}{P_i}\right) \, .
\ee
where the $I_i$ are the NS moments of inertia.
Label the faster-rotating NS ``1''.  Assuming $\chi_1 >> \chi_2$, the precession
amplitude is simply
\beq
\begin{split}
\lambda_{\rm L} \approx 2.3 \times 10^{-4}& (1 - {\rm cos}^2\theta_{\rm LS})^{1/2}\\
&\cdot\left(\frac{\chi_1}{0.01}\right) \left(\frac{M(1+z)}{2.8 M_{\odot}}\right)
\left(\frac{f}{0.3\,\rm Hz}\right)^{1/3} \, .
\end{split}
\eeq
where $\theta_{\rm LS}$ is the angle between $\vec L$ and $\vec S_1$.   
If we ignored spin-orbit precession when subtracting out the NS inspiral waveforms, we would make
relative errors $\delta h/h \sim \lambda_{\rm L}$.  This is $ \lesssim 10^{-3}$ for $P_1 \gtrsim 10\,$ms, 
and so these errors would typically be benign.
In any cases where $P_1$ is significantly less than $10$\,ms, 
this will generally be clear from the data (from its influence on the orbital phase evolution) and
these very-high-spin systems would presumably be treated as a ``special class'', 
requiring more parameters to fit them than typically necessary.

\subsubsection{Effect of spin-orbit and spin-spin terms on waveform phase}
We next consider the effect of the spin-orbit and spin-spin interactions on
the waveform phase. Since we have considered the effects of orbital eccentricity and 
orbital-plane precession in previous subsections, we simplify the analysis here by 
assuming that the orbit is circular and that the orbital angular momentum vector $\vec L$ and the
two spin vectors, $\vec S_1$ and  $\vec S_2$, are all aligned.
Then in a post-Newtonian expansion of the waveform phase $\Psi(f)$, 
the lowest order terms involving the spin-orbit and spin-spin interaction are~\cite{Vec2004}
\be
\begin{split}
\Psi_{\beta}(f) + \Psi_{\sigma}(f) = 
\frac{3}{4}&\left(8 \pi {\cal M} (1+z) f \right)^{-5/3} \\
&\times\left[4\beta\, y^{3/2} - 10 \sigma\, y^2 \right].
\end{split}
\ee
\noindent
where the terms $\beta$ and $\gamma$ are explicitly given by
\beq
\label{beta}
\begin{split}
\beta \equiv & \left(\frac{113}{12} +\frac{25}{4}\frac{M_2}{M_1}\right)(M_1/M)^2(\hat L \cdot \hat S_1)\chi_1 \\
&+ \left(\frac{113}{12} +\frac{25}{4}\frac{M_1}{M_2} \right)(M_2/M)^2(\hat L \cdot \hat S_2)\chi_2
\end{split}
\eeq
\noindent and
\be
\label{sigma}
\sigma \equiv \frac{\mu}{M} \chi_1 \chi_2 \left(\frac{247}{192} \hat S_1
\cdot \hat S_2 - \frac{721}{192}(\hat L \cdot \hat S_1)(\hat L \cdot \hat S_2)
\right)\, .
\ee
Assuming $P_1 \sim 30\,$ms and $P_2 \sim 1\,$s, this implies
$\chi_1 \sim 0.01$ and $\chi_2 \sim 4\times 10^{-4}$, and then
$\beta \sim 0.04$, while $|\sigma| \sim 2.5 \times 10^{-6}$.
So plugging in fiducial values (with $M_1 = M_2 = 1.4 M_{\odot}$), the spin-related phase terms are
\beq
\begin{split}
\Psi_{\beta}(f) & \sim  6.8 \times 10^1 \left(\frac{\beta}{0.1}\right)\left(\frac{f}{0.3 \rm Hz}\right)^{-2/3} (1+z)^{-2/3} \\
\Psi_{\sigma}(f) & \sim  -4 \times 10^{-4} \left(\frac{\sigma}{10^{-5}}\right)\left(\frac{f}{0.3 \rm Hz}\right)^{-1/3} (1+z)^{-1/3} 
\end{split}
\eeq

In summary, the spin-orbit term $\beta$ is clearly relevant, 
while spin-spin term $\sigma$ is negligible for typical cases.
Thus, while it takes $6$ parameters to describe (initial conditions for)
the two spin vectors $\vec S_1$ and $\vec S_2$, for typical cases
the spins' influence on the waveform can be adequately 
subsumed into a single parameter, $\beta$.

\subsection{High-Order post-Newtonian Effects, neglecting spin}
\label{secPN}
To-date, the post-Newtonian equations governing the inspiral
of (quasi-)circular-orbit
binaries have been derived through $\rm P^{3.5}N$ order beyond the
lowest-order, quadrupole-formula level~\cite{Nissanke_Blanchet_05}. 
 Is that good enough for
accurately subtracting out the merger waveforms from the BBO data, or
are even higher-order treatments called for? In this subsection, we
do a rough estimate that suggests that the $\rm P^{3.5}N$ equations are
sufficiently accurate for this purpose (or are at least very close).  
Since we have considered the effects of spin and orbital eccentricity in previous subsections, for this
subsection we will specialize to the case of nonspinning NSs in (quasi-)circular orbits.

We return again to the stationary-phase approximation of the waveform
\be
\tilde h(f) \propto \big({\cal M}(1+z)\big)^{5/6}f^{-7/6}[1 + \ldots] \e^{\irm \Psi(f)}
\ee
and to the PN expansion of the phase $\Psi(f)$:
\beq
\label{eqpNpsi}
\begin{split}
\Psi(f) = &
\;{\rm const} + 2\pi f t_{\rm c} +\frac{3}{4}\left(8 \pi {\cal M} (1+z) f \right)^{-5/3} \\
\mbox{} &  \times \, \left[1 + \frac{20}{9}\left(\frac{743}{336}+\frac{11\mu}{4M}\right) y -16 \pi y^{3/2} + \ldots \right].
\end{split}
\eeq
Terms up through $\rm P^{3.5}N$ have already been calculated.
We want to estimate the size of the $\rm P^4N$ term in the series, which
corresponds to a term of the form
$\frac{3}{4}\big(8 \pi {\cal M}(1+z) f )^{-5/3} \times \big[\big(C + D(\mu/M) + E(\mu/M)^2 + \cdots\big) y^4\big]$,
for some coefficients $C, D, E, \cdots$.   The coefficient C could be derived from the
results in \cite{SaTa2003}; we have not done that calculation, but
it is clear from \cite{SaTa2003} that C is of order $10^2$.  It seems reasonable
to assume that the sum $C + D(\mu/M) + E(\mu/M)^2 + \cdots$  
is also $\sim 10^2$. The {\it rest} of the
$\rm P^4N$ term,  $\frac{3}{4}\big(8 \pi {\cal M}(1+z) \, f)^{-5/3} \, y^4$, has magnitude
\be
4.06 \times 10^{-6} \left(\frac{M(1+z)}{2.8 M_{\odot}}\right)\left(\frac{f}{1\,{\rm Hz}}\right)
\ee
and so the full term is of order $10^{-3}$ at $f= 1\,$Hz. 

Thus the $\rm P^4N$ contribution is
just at the border of being relevant. We suspect the full $\rm P^4N$ term will have been 
calculated long before BBO flies, but even today one could generate a ``poor man's'' $\rm P^4N$ waveform by simply omitting the terms involving $D(\mu/M)$, $E(\mu/M)^2$, etc., 
but including the  term $\propto C$, which we repeat is easily derivable from published results.
Because $\mu/M \approx 1/4$, the omitted terms could easily be an order of magnitude smaller
than the C-term, and so would be truly negligible.

Therefore we believe that already, today, one could produce PN waveforms 
that are sufficiently accurate for BBO, or that are at least quite close. 
However we add that if this view turned out to be too optimistic--
if it {\it did} prove difficult
to generate sufficiently accurate waveforms, corresponding to realistic solutions of Einstein's
equation--then there is also an obvious fall-back strategy: use an enlarged space of ``phenomenological
waveforms,'' such as those developed by Buonanno et al.
~\cite{BuEA2004},
to identify and subtract out the inspirals. 
The family of phenomenological waveforms would depend on a few more parameters than the
physical waveforms, so projecting out subtraction errors would cost somewhat more bandwidth, but
the estimates in Sec.~IV.B show that this cost would still likely be minimal. 
Therefore as long as {\it some} member of the phenomenological family lies quite close
to each true waveform, meaning $\delta h/h \lesssim 10^{-3}$, the phenomenological family would suffice for the
purposes of inspiral-waveform subtraction.

\section{The Detection Threshold $\rho_{\rm th}$}
The GW strength (at the Earth) of any NS-NS binary is
characterized by its signal-to-noise-squared, $\rho^2 $.
By $\rho^2$, we mean the matched-filtering SNR$^2$
for the entire 4-constellation BBO network (whose output
is $12$ independent GW data streams, $8$ of which 
have good sensitivity to NS binaries).
We want to estimate the threshold value $\rho_{\rm th}^2 $ required for the
signal to be detectable. There are basically two sorts of considerations here.
If one possessed infinite computing power, then this threshold value
would be set just by the requirement that one has sufficient confidence
in the detection (i.e., that the false alarm rate be sufficiently low).
However in practice we expect the search sensitivity to be (severely) 
computationally limited, which implies a somewhat higher detection
threshold.

\subsection{Lower bound on $\rho_{\rm th}$ set by the number of effectively independent
inspiral templates}

Let $N_{\rm t}$ be the number of independent templates
required to cover the parameter space of NS-NS inspiral waveforms
('independent' in the sense that 
they have only modest overlap with each other).
Then for a given threshold value
$\rho_{\rm th}$, the number of false alarms generated by this entire
set is $\sim N_{\rm t} \,{{\rm erfc}}(\rho_{\rm th}/\sqrt{2})
\approx N_{\rm t}\,(2\pi)^{-1/2} (\rho_{\rm th})^{-1} \e^{-\rho^2_{\rm th}/2}$.
In practice,
one would probably want this false alarm rate to be no greater than
$\sim 0.01$. How large is $N_{\rm t}$ for our problem?
This has not yet been calculated, but because $\rho_{\rm th}$  depends only
logarithmically on $N_{\rm t}$, a very rough estimate will suffice for our purposes.

Consider the parameter space of 'typical' inspiral waveforms, normalized by
$\langle {\bf h} \, | \, {\bf h}\rangle = 1$.  
These are effectively described by 10 parameters:
\beq \label{lambda}
\begin{split}
\lambda^\alpha &\equiv (\lambda^1,\ldots,\lambda^{\rm N}) \\
&=
\left[t_0,\, \ln{\cal M}_{\rm eff} ,\,\ln\mu_{\rm eff}, \,\beta,\, e^2_0,\, \Phi_0,\,
\theta,\, \phi,\, \theta_{\rm L},\, \phi_{\rm L} \right].
\end{split}
\eeq
Here, $t_0$ is the instant of
time when the ($n=2$ piece of the) GW frequency sweeps through
some fiducial value $f_0$ (e.g., $f_0 = 0.3\,$Hz); ${\cal M}_{\rm eff} \equiv {\cal M}(1+z)$ 
$\mu_{\rm eff} \equiv \mu(1+z)$; 
 $\beta$ is the spin parameter 
defined in Eq.~(\ref{beta}) (and approximated here
as a constant);
$e^2_0$ is the square of the orbital eccentricity at $t_0$;
$\Phi_0$ describes the orbital phase (the angle between the
orbital separation vector $\hat r$ and some fixed vector in the orbital
plane) at $t_0$;
$(\theta,\, \phi)$ give the position of the source on the sky; and
$(\theta_{\rm L}, \phi_{\rm L})$ give the orientation of the binary's total
angular momentum vector $\vec L$ (which precesses slightly, but which 
we can typically approximate as constant).
We have omitted from this list the perihelion angle $\omega_0$ 
and 5 of the 6 parameters characterizing the two NS spin vectors, since
we estimated in Sec.~V.C that they typically have a negligible impact on the 
waveform.  The luminosity-distance to the source, $D_{\rm L}$, 
has been omitted since it affects only the waveform's overall normalization.

Now imagine covering our N-dimensional manifold of waveforms
with a hypercubic grid, such that
the overlap of any waveform on the manifold with the nearest
gridpoint is $\ge (1-x)$,  where $x$ is a number that characterizes the
fineness of our grid. The number of
gridpoints $N_{\rm t}$ is then~\cite{Owe1996}
\be
N_{\rm t} \approx (N/8x)^{N/2}
\int{\sqrt{\Gamma} \,\drm\lambda^1 \ldots  \drm\lambda^{\rm N}}
\ee
where $\Gamma$ is the determinant of the Fisher matrix 
$\Gamma_{\alpha\beta} \equiv \langle \partial_{\alpha} {\bf h} \, | \,\partial_{\beta} {\bf h}\rangle$
(again, subject to the constraint $\langle {\bf h} \, | \, {\bf h}\rangle = 1$).
In our case $N=10$, and we adopt $x = 0.5$ as our fiducial grid spacing, so 
$(N/8x)^{N/2} \approx 100$.
We can obtain a rough estimate of the integral
$\int{\sqrt{\Gamma} \drm\lambda^1 \ldots  \drm\lambda^{\rm N}}$
from estimates of the sizes of the
diagonal elements of $\Gamma$, as follows.
For each parameter $\lambda_{\alpha}$, let $n_{\lambda_{\alpha}} = \delta_{\lambda_{\alpha}}
|h^{-1}\partial h/\partial{\lambda_{\alpha}}|$, where  $\delta_{\lambda_{\alpha}}$
is the range of integration for the ${\alpha}^{\rm th}$ parameter and
$|h^{-1}\partial h/\partial{\lambda_{\alpha}}|$ is supposed to represent some
'typical' or 'rms' value of this quantity. Then
\be
\int{\sqrt{\Gamma}\, \drm\lambda^1 \, \drm\lambda^2 \ldots  \drm\lambda^{\rm N}} \lesssim
n_{\lambda_1}\,  n_{\lambda_2}  \ldots  n_{\lambda_{\rm N}} \, .
\ee
The rhs represents a rough upper limit to the integral because it
ignores possible cancellations in the determinant coming from
the off-diagonal terms.
Based on a post-Newtonian expansion of the waveform, of the form shown
above in \S \ref{secPN}, we derive the following order-of-magnitude estimates for the
different factors:

\bea\label{pn_estimate}
&&n_{t_0} \sim 10^8,\,n_{\ln{\cal M}} \sim 10^8,\,n_{\ln\mu} \sim 10^5, \nonumber\\
&&n_{\beta} \sim 10^2,\,n_{e^2_0}\sim 10^2, \,n_{\Phi_0} \sim 10^1,\,n_{\Omega} \sim 10^7, \nonumber \\
&&n_{\Omega_{\rm J}} \sim 10^1
\eea
\noindent where $n_{\Omega} \equiv n_{\theta} n_{\phi}$,
$n_{\Omega_{\rm J}} \equiv n_{\theta_{\rm J}} n_{\phi_{\rm J}}$, and
where we have used $\delta \beta \sim 0.5$ and $\delta e^2_0 \sim 10^{-7}$.
Using the above estimates, we find $N_{\rm t} \lesssim 10^{36}$.
Allowing for cancellations from off-diagonal terms, it seems
reasonable to assume  $N_{\rm t}$ is in the range $N_{\rm t} \sim 10^{30}-10^{36}$,
implying $\rho_{\rm th} \ge 12.5 - 13.5$. That is, if matched filtering
reveals a NS-NS inspiral with total ${\rm SNR} \gtrsim 13$, then one can be confident
it is not simply a randomly generated peak. 

Now, one could complain that we have undercounted $N_{\rm t}$ by restricting to 
the parameter space of 'typical' signals, whereas among the
$10^5-10^6$ NS binaries that BBO will observe, there are probably {\it some}
atypical ones; e.g., binaries in which {\it both} NSs are
rapidly rotating.  And these must also be identified and subtracted, for BBO to 
do its main job. This complaint has some merit, but we do not dwell on it here,
since in any case we expect that in practice $\rho_{\rm th}$ will be set not 
by the false alarm rate, but by
computational limitations. We turn to these next.

\subsection{Limitations due to Finite Computing Power}
>From the estimates in the previous section, one readily concludes that straightforward matched
filtering for all templates in the template bank will {\it not} be possible.
The simplest implementation would require of order $ \sim 10^9\, N_{\rm t}$
floating point operations (since each year-long template has $\sim 3\times 10^8$ data points, if
sampled at $\sim 10\,$Hz). A well known, FFT-based trick to efficiently search over
all $t_0$~\cite{Schutz} reduces this cost by a factor $\sim n_{t_0}/[3 \ln(10^9)] \sim 10^6$, 
but would still require
computation speeds of $\sim 10^{28 \pm 3}$ flops (operations per second).
Extrapolation of Moore's law to the year 2025
suggests that perhaps $\sim 10^{17}$ flops will be readily available, which
is 11 orders of magnitude too small for the job.

Therefore one will need to devise a suboptimal (but
computationally practical) search algorithm, and live with the
attendant loss in sensitivity. It is beyond the scope of this paper
to design such an algorithm and evaluate its efficiency.
Fortunately, though, the problem of searching for NS-NS binary signals
in BBO data is closely analogous to the problem of searching for
unknown GW pulsars in LIGO data, and the problem of devising
efficient search algorithms
for the latter has been studied in some detail~\cite{BrCr2000,CGK2005}.
We will estimate the threshold sensitivity of BBO NS-binary searches
based on this analogy, so we digress to describe optimized
LIGO searches for unknown GW pulsars.

By unknown GW pulsars, we mean rapidly rotating NSs whose
sky location, amplitude and polarization, and gravitational-wave frequency
(at any instant) and frequency derivatives are all unknown, and so must
be searched over. I.e., the unknown parameters are 
the sky location ($\theta, \phi$), four parameters describing
the amplitude, polarization, and overall phase of the waves
(these can be usefully thought of as two complex amplitudes--one for each
GW polarization),
and the gravitational wave frequency and
frequency derivatives at any instant:
$f, \dot f$, $\ddot f$, $\dddot f$, etc.
The typical magnitude of frequency derivatives is assumed to be
$\drm^n f/\drm t^n \sim f/\tau^n$, where $\tau$ is some characteristic
timescale (basically the NS's spindown-age), but these derivatives
are otherwise considered independent.

For GW pulsars, we briefly describe the most efficient schemes that have been considered to-date,
which are semi-coherent and hierarchical (i.e., multi-stage) searches;
we refer to Cutler et al.~\cite{CGK2005} for more details.
A ``semi-coherent''  search is one where
short data stretches (say, a few days long) are all coherently searched, using
some technique akin to matched filtering, and then the resulting
powers from the different stretches are summed.  The method is only
``semi-coherent'' because powers are added instead of complex
amplitudes; i.e., information regarding the overall phase of the
signal in different stretches is discarded. This allows one to use a
much coarser grid on parameter space than would be required in a fully
coherent search of the same data.
The basic idea of multi-stage searches is as follows. In the first
stage one searches, semi-coherently, through some fraction of the
data (say, a month's worth), and identifies promising
``candidates'' in parameter space. One then follows up these candidates
in the second stage, using a higher resolution on parameter space (a finer
grid) and more data. This generates a second, sublist of candidates, which
one then investigates with even higher resolution and yet more data, and so on.
The idea is to reject unpromising regions in parameter space as quickly
as possible, so as not to waste valuable computer resources on them.
After $N_{\rm s}$ semi-coherent stages like this, any remaining candidates are
verified using a final, fully coherent follow-up search in a very
tiny region of parameter space. A priori, the best value for $N_{\rm s}$ is unclear;
it was shown in Cutler et al.~\cite{CGK2005} 
that for realistic GW pulsar searches,
the gains from increasing $N_{\rm s}$ saturate
at $N_{\rm s}=3$ semi-coherent stages.

The GW signal from a NS binary is practically the same as the signal from
a low-frequency GW pulsar (except the binary's orbital frequency changes on a much shorter
timescale than the spin-period of slowly rotating NSs).
In both cases, the signal is essentially monochromatic at any instant, with
a frequency that is slowly time-varying.
In both cases there is an unknown sky position, two unknown complex
amplitudes (equivalent to $D$, $\theta_{\rm L}$, $\phi_{\rm L}$, and $\Phi_0$ in the
NS-binary case).
The optimal statistic for searching over the two complex (four real)
amplitudes, in both the GW-pulsar and NS-binary cases,
is the F-statistic, which follows a chi-squared distribution
with 4 degrees of freedom~\cite{BPD2000,CuSc2005}. (The distribution
is the same no matter how many detectors are combined in the analysis;
the 4 d.o.f. correspond to the 2 complex--or 4 real --unknown
amplitude parameters.
The fact that BBO is composed of 4 LISA-like constellations outputting
12 independent data streams does not affect this counting.)
The biggest difference between the two sources is that for actual GW pulsars, the signal's
intrinsic amplitude can be approximated as constant over the observation time,
while in the NS-binary case, the GW amplitude grows significantly during
the observation time. However we do not consider this difference as very important 
when comparing detection thresholds, 
especially because the search sensitivity is really set by the early stages,
where the coherent integration times will be significantly shorter than
one year.

The sensitivity of the GW-pulsar search is limited by the size of the parameter
space one wishes to search; e.g., for an all-sky search, the size of the
parameter space is set by the maximum frequency $f_{\rm max}$ and the
shortest spin-down age $\tau_{\rm min}$ that one wishes to search over.
We now try to choose a search-space that makes the LIGO GW pulsar
search comparable in difficulty to the BBO NS-binary search.
The pulsar parameters ($\dot f$, $\ddot f$, $\dddot f$) are
closely analogous to the NS-binary parameters $(\cal M, \mu, \beta)$,
which control the inspiral rate. Assuming a search up to frequency
$f_{\rm max} = 1000$\, Hz, and an observation time of $T_0 = 1\,$yr,  we
estimate $n_{\dot f} n_{\ddot f} n_{\dddot f}  \sim
(f_{\rm max} T_0)^3 (T_0/\tau_{\rm min})^6 \sim 3\times 10^{31} (1\, {\rm yr}/\tau_{\rm min})^6$. 
Using the estimates from Eq.~(\ref{pn_estimate}), we find that
$n_{\dot f} n_{\ddot f} n_{\dddot f}  \sim
n_{\ln{\cal M}} n_{\ln\mu} n_{\beta}$ for $\tau_{\rm min} \sim 300\,$yr.

Continuing our comparison of the LIGO/pulsar and BBO/binary searches, we note that because three
of BBO's four constellations have separations of order $1\,$AU ($\approx 500$\,s),
the number of distinct patches on the
sky that must be searched over is 
$\sim (4\pi) (2\pi \times 0.3\,{\rm Hz} \times 500\,{\rm s})^2 \sim 10^7$.
In comparison, for GW pulsar searches, the number of distinct sky patches
is set by the Earth's rotation about its axis,
and is $\sim 3 \times 10^4$,  
or roughly 300 times fewer.  (This counting assumes that the larger, but more slowly
varying, Doppler shift due to
Earth's motion around the Sun can be absorbed into the unknown pulsar
spin-down parameters, which should be true for integration times shorter than
a few months. This is good enough for our purposes, since the sensitivity
of the search is really set at early stages, where only a month or two of data
is examined.)
On the other hand, assuming sampling at $\sim 10\,\rm Hz$ for
BBO and sampling at $\sim 3\,\rm kHz$ for the LIGO network, a year-long
GW pulsar template contains $\sim 300$ times as many points as a
year-long BBO NS-binary template, so each coherent integration
requires about 300 times more floating point operations in the
LIGO/pulsar case than in the BBO/binary case.

Therefore we conclude that a LIGO/pulsar search for unknown NSs, over a parameter range
set by $(f_{\rm max}= 1000\,{\rm Hz}, \tau_{\rm min} = 300\,{\rm yr})$, 
is comparable in difficulty, computationally, 
to the BBO/binary search.
The code used by Cutler et al.~\cite{CGK2005} to calculate the efficiencies of 
multi-stage GW pulsar searches was re-run for this parameter range, assuming an available computational 
power of $10^{17}$ flops (and computation time of one year).
For this parameter range and computational power, LIGO/pulsar search with 3 semi-coherent
stages (plus a final, coherent follow-up) should be able to detect GW
pulsars with $\rho$ as small as $20$~\footnote{I.~Gholami, private communication} (with false-dismissal rate = $10\%$ and false-alarm rate = $1\%$). 
Therefore we estimate that 
BBO will {\it also} be able to detect and remove
NS binaries with $\rho > \rho_{\rm th} = 20$
(or roughly $50\%$ higher than the minimum $\rho_{\rm th} \sim 13$ required for detection confidence).  

However: as in the last subsection, one could complain that we have counted only the cost of
searching for 'typical' binaries, whereas in practice most of the computational budget
may be spent on searching for the few atypical ones.  
Also, we have assumed (reasonably, we think, but without justification) 
that the computational cost
of identifying all the individual sources is greater than (or at least comparable to) 
the cost of finding the combined best fit. 
Also, the comparison was made for a single false-dismissal rate ($10\%$), 
whereas we imagine that, in actual practice for the BBO analysis, one would want
to do the BBO analysis in stages, with an ever-decreasing FD rate.
Also, actual BBO searches may be plagued by many more 
outliers than would be present for the purely Gaussian noise
that our sensitivity estimates were based on, and this would increase the threshold.
For all these reasons, 
and because our method of estimating $\rho_{\rm th} \approx 20$ ``by analogy'' was so crude 
in the first place,
we will investigate the efficacy of NS-binary subtraction 
for a range of detection thresholds: $\rho_{\rm th} = 20$, $30$, or $40$.

\section{Equations characterizing a self-consistent subtraction scheme}
Fix the values of the merger rate $\dot n_0$
(which sets the overall magnitude of $S_{\rm h}^{\rm NSm}$)  and the detection threshold $\rho^2_{\rm th}$.
We want to calculate what fraction $F^2$ of the spectral density of the NS-binary foreground
{\it cannot} be subtracted. 
For simplicity, we will assume that all NSs have mass $1.4 M_\odot$
Then our method for self-consistently 
determining $F^2$ proceeds by the following steps.

Step 1: Adopt some initial ``guess'' value $F^2_{\rm G}$. Based on this guess, we 
obtain a corresponding guess for the total noise level:
\beq\label{noise-guess}
S^{\rm tot}_{\rm h}(F_{\rm G}, \, f)=S_{\rm h}^{\rm inst}(f) +F_{\rm G}^2\cdot S_{\rm h}^{\rm NSm}(f)
\eeq

Step 2: Based on this total noise level, we determine the redshift $\bar z(\mu)$, out to which
a NS-binary with orientation $\mu \equiv \hat L \cdot \hat N$ can be detected.
This boundary $\bar z(\mu)$ (separating detectable and undetectable sources) 
is determined by the equation (derived in the Appendix):

\beq
\rho_{\rm th}^2 =8\cdot\frac{2f(\mu)}{3\pi^{4/3}}\frac{(\mathcal{M}(1+\bar{z}))^{5/3}}{D_L^2(\bar{z}) }\int\limits_0^\infty\drm f\,\frac{f^{-7/3}}{S_{\rm h}^{\rm tot}(F_{\rm G}, \, f)}.
\label{eqSN}
\eeq
where the function $f(\mu)$ gives the dependence of the squared waveform amplitude on $\mu$:
\beq f(\mu)\equiv
\frac{(1+\mu^2)^2+4\mu^2}{\int\limits_0^1\drm\mu\,\left[(1+\mu^2)^2+4\mu^2\right]}=\frac{5}{16}(\mu^4+6\mu^2+1).
\eeq
Note we have normalized $f(\mu)$ so that $\int_0^{1} f(\mu) d\mu = 1$.  

Step 4: We compute the fraction $F^2$ of the NS-binary foreground that is due to sources more
distant than $\bar z(\mu)$. Based on Eqs.~(10) and (12) in Phinney~\cite{Phi2003}, this fraction
is easily seen to be 

\beq
F^2 = \frac{S_{\rm h}^{{\rm NSm}, >\bar{z}}(f)}{S_{\rm h}^{\rm NSm}} = 
\,\frac{1}{C(0)}\int\limits_0^1\drm\mu\ f(\mu) C\big(\bar z(\mu)\big) \, ,
\label{eqNSNSfg}
\eeq
in terms of the integral
\beq
C(\bar z)\equiv\int\limits_{\bar z}^\infty\drm z\frac{r(z)}{(1+z)^{4/3}H(z)}.
\eeq
where $H(z)$ and $r(z) \equiv \dot n(z)/\dot n_0$ are given explicitly in 
Eqs.~(\ref{Hz}) and (\ref{rz}), respectively.

So far, we have given an algorithm for computing $F(F_{\rm G})$, i.e., for iteratively
improving our initial guess $F_{\rm G}$. An initial guess $F_{\rm G}$ leads to a 
self-consistent solution if $F(F_{\rm G}) = F_{\rm G}$. Clearly, we can short-cut the iterative procedure
simply by looking for fixed points of this function. I.e., our last step is 

Step 5: Plot $F(F_{\rm G})$, and look for fixed points, i.e., values $F_{\rm G}$ such that
$F(F_{\rm G}) - F_{\rm G} = 0$.  

Our results are displayed in the next section.

\section{Results}
As motivated in previous sections, we calculate the efficacy of foreground subtraction for $3$ different values of the present-day merger rate density, 
$\dot{n}_0=\{10^{-8},\,10^{-7},\,10^{-6}\}\,{\rm yr}^{-1}{\rm Mpc}^{-3}$, and $3$ values of the detection threshold, $\rho_{\rm th}=\{20,\,30,\,40\}$.
This yields 9 different results for the self-consistent $F$ representing the fraction of the foreground noise amplitude due to undetectable
(and hence unsubtractable) NS binaries.
 We calculate these results both for the 
``standard BBO'' design sensitivity, $S_{\rm h}^{\rm st. inst}(f)$, shown in Fig.~2, and for a less sensitive version having 
$S_{\rm h}^{\rm inst} = 4\times S_{\rm h}^{\rm st. inst}(f)$, 
i.e., with $2\times$ higher instrumental noise {\it amplitude}.  As a shorthand, we will refer to the latter as 
``standard/2'' sensitivity. Our main results are presented in Sec.~VIII.A.  
In Sec.VIII.B  we gain insight into our results by exploring {\it which} binaries (i.e., which $z$ and $\mu$) are undetectable, for
different $\dot n_0$ and $\rho_{\rm th}$.
Finally, in VIII.C, we consider the case of a larger foreground, 
$\dot{n}_0=10^{-5}\,{\rm yr}^{-1}{\rm Mpc}^{-3}$; although this merger rate is
unrealistically high, this case provides a rather
interesting illustration of our general method. 

\subsection{Efficacy of Background Subtraction for BBO with Standard and Standard/2 Sensitivity}
In Sec.~VII we showed that self-consistent F values are 
fixed points of the 
function $F(F_{\rm G})$, where $F_{\rm G}$ denotes
a ''guessed'' value for this fraction.  
For standard BBO sensitivity, we find that the solution $F$ is practically independent of $\dot n_0$, for 
realistic merger rates.
Specifically, we find
\beq
\begin{split}
F_{20} &= F_{30} = 0,\\
F_{40} &= 0.0015 \, ,
\end{split}
\label{eqFBBO}
\eeq
where our notation is that $F_{20}$ 
is the solution $F$ for $\rho_{\rm th} = 20$, assuming the standard BBO instrumental noise level, and 
similarly for $F_{30}$  and $F_{40}$.  
Therefore standard BBO is sensitive enough that the NS-NS foreground can be entirely (or almost entirely) subtracted,
independent of the merger rate or detection threshold (for realistic values of those quantities).

\begin{figure}[ht!]
\centerline{\includegraphics[width=9.0cm]{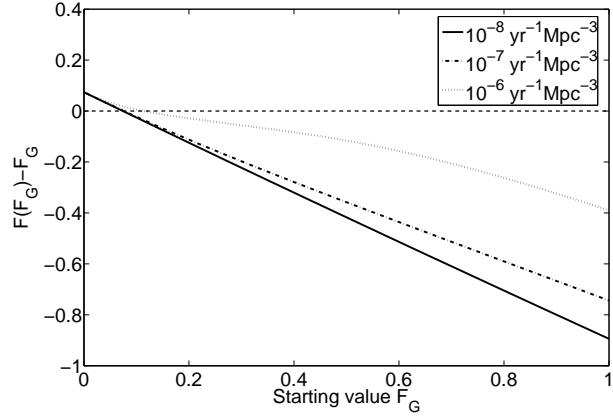}}
\caption[Binary subtraction (increased $S_{\rm h}^{\rm inst}$)]{Shows the function $F(F_{\rm G})-F_{\rm G}$ 
for three merger rates: $\dot{n}_0=\{10^{-8},\,10^{-7},\,10^{-6}\}\,{\rm yr}^{-1}{\rm Mpc}^{-3}$. 
All curves are for ``standard/2'' sensitivity and detection threshold $\rho_{\rm th}=30$. }
\label{figFixD}
\end{figure}

Next we consider BBO with ``standard/2'' sensitivity. 
We denote by  $F_{20}^{\prime}$ 
the self-consistent solution for $\rho_{\rm th} = 20$ and standard/2 sensitivity, 
and similarly for $F_{30}^{\prime}$  and $F^{\prime}_{40}$.  For this case,  the results 
do generally depend on $\dot n_0$ (unlike for standard BBO).
Our nine results for $F^{\prime}$, corresponding to the nine combinations of $(\dot n_0, \rho_{\rm th})$, are 
given in Table~II.  To illustrate how these results are derived, in Fig.~\ref{figFixD} we show the 
function $F(F_{\rm G})-F_{\rm G}$ for each $\dot n_0$, and for fixed $\rho_{\rm th} = 30$. 
The entries in the second row of Table~II are just the 
the $F_{\rm G}$ values where the three curves in Fig.~\ref{figFixD} 
pass through zero.

\renewcommand{\arraystretch}{1.7}
\setlength{\tabcolsep}{3mm}
\begin{center}
\begin{table}[ht]
\begin{tabular}{|c||c|c|c|}
\hline
& \multicolumn{3}{|c|}{$S_{\rm h}^{\rm inst}=4\cdot S_{\rm h}^{\rm st.inst}$} \\
\cline{2-4}
\raisebox{-0.5ex}{$\rho_{\rm th}$}$\;\;\;$\raisebox{1ex}{$\dot{n}_0$} & $10^{-8}$ & $10^{-7}$ & $10^{-6}$ \\
\hline\hline
20 & 0.0015 & 0.0015 & 0.0015 \\
\hline
30 & 0.071 & 0.077 & 0.11\\
\hline
40 & 0.15 & 0.17 & 0.55\\
\hline
\end{tabular}
\caption{Results for ``standard/2'' sensitivity.  Table lists $F^\prime=(S_{\rm h}^{{\rm NSm}, >\bar{z}}/S^{\rm NSm}_{\rm h})^{1/2}$
for different combinations $(\dot n_0, \rho_{\rm th})$.  $F^\prime$ is the
amplitude of confusion noise from unsubtractable NS binaries, divided by the total foreground amplitude.} 
\label{tabFinc}
\end{table}
\end{center}

None of the $F^{\prime} $ values is Table~II is zero;  which ones are sufficiently small that 
unsubtracted binaries would not significantly interfere with BBO's main goal?
To answer this, in Table~\ref{tabSuccess} we give the ratio
$[S_{\rm h}^{{\rm NSm},>\bar{z}}(f)/S^{\rm GW}_{\rm h}(f)]^{1/2}$, evaluated at $f=1\,$Hz, 
for each combination $(\dot n_0, \rho_{\rm th})$.  Again, 
$S_{\rm h}^{{\rm NSm},>\bar{z}}(f) \equiv (F^{\prime})^2 S_{\rm h}^{\rm NSm}(f)$, while 
in Table~III 
$[S^{\rm GW}_{\rm h}(f)]^{1/2}$ is the noise spectrum for a primordial background with
$\Omega_{\rm GW}(f) = 10^{-15}$.
Thus, ratios smaller than one indicate that the 
unsubtracted piece of the foreground is smaller than a primordial background with this
energy density.
We see that if $\rho_{\rm th} = 20$ (i.e., if the detection pipeline can uncover almost all NS binaries with
total $\rm SNR = 20$), then even with standard/2 sensitivity, BBO would still 
be able to detect a primordial background having $\Omega_{\rm GW}(f) \ge 10^{-15}$.

However Table~\ref{tabSuccess} also shows that if $\rho_{\rm th} = 30$ or $40$, 
and instrumental sensitivity is standard/2, then
BBO would be unable to detect primordial background of $\Omega_{\rm GW}(f) \sim 10^{-15}$
(since it would be ``covered up'' by the unsubtractable part of the foreground).

We point out that entries for the case $(\dot n_0 = 10^{-8}, \rho_{\rm th} = 20)$ 
in Tables II and III
should not be taken too literally, since in that case our solution $F^\prime$ corresponds to less than 
one unsubtracted binary.  (A single merging NS binary at $z = 5$, even with $\mu = 0$, contributes
$\sim 10^{-18}$ to our local $\Omega_{\rm GW}(f)$, in the BBO band.)
What this means, of course,
is that our solution $F^\prime$ lies outside the range of validity of our equations, whose
derivation implicitly assumed that at least one source was undetectable. Just as clearly, 
our main conclusions are unaffected.
The proper interpretation of the $(\dot n_0 = 10^{-8}, \rho_{\rm th} = 20)$ entries is that, for these
values, BBO with standard/2 sensitivity would likely detect every single 
NS-NS merger occurring on its past light cone.  
\begin{center}
\begin{table}[ht]
\begin{tabular}{|c||c|c|c|}
\hline
& \multicolumn{3}{|c|}{$S_{\rm h}^{\rm inst}=4\cdot S_{\rm h}^{\rm st.inst}$} \\
\cline{2-4}
\raisebox{-0.5ex}{$\rho_{\rm th}$}$\;\;\;$\raisebox{1ex}{$\dot{n}_0$} & $10^{-8}$ & $10^{-7}$ & $10^{-6}$ \\
\hline\hline
20 & 0.030 & 0.10 & 0.30 \\
\hline
30 & 1.4 & 4.9 & 22\\
\hline
40 & 3.0 & 11 & 110\\
\hline
\end{tabular}
\caption{Table of ratios $[S_{\rm h}^{{\rm NSm}, >\bar{z}}(f)/S^{\rm GW}_{\rm h}(f)]^{1/2}$ evaluated at $f=1\,$Hz, for 
BBO with standard/2 sensitivity.  Here $S^{\rm GW}_{\rm h}(f)$ is from  
a primordial background with $\Omega_{\rm GW}(f) = 10^{-15}$.  Ratios smaller than one 
indicate that the unsubtractable part of the NSm foreground noise is smaller than this primordial background level.
The results here are equivalent to those in Table~II.}
\label{tabSuccess}
\end{table}
\end{center}

We also repeated the above analysis for BBO with only standard/4 sensitivity, i.e, with $S_{\rm h}^{\rm inst}(f) = 16\cdot S_{\rm h}^{\rm st.inst}(f)$.
This noise level is clearly inadequate, since even for $\rho_{\rm th} = 20$ and a low merger rate, 
$\dot n_0 = 10^{-8}\,{\rm yr}^{-1}{\rm Mpc}^{-3}$, we find
$[S_{\rm h}^{{\rm NSm}, >\bar{z}}(f)/S^{\rm GW}_{\rm h}(f)]^{1/2} \approx 3.0$ at 
$f=1\,$Hz, for $\Omega_{GW}(f) = 10^{-15}$.

\subsection{Further analyses of the subtraction scheme}
\begin{figure}[ht]
\centerline{\includegraphics[width=9.0cm]{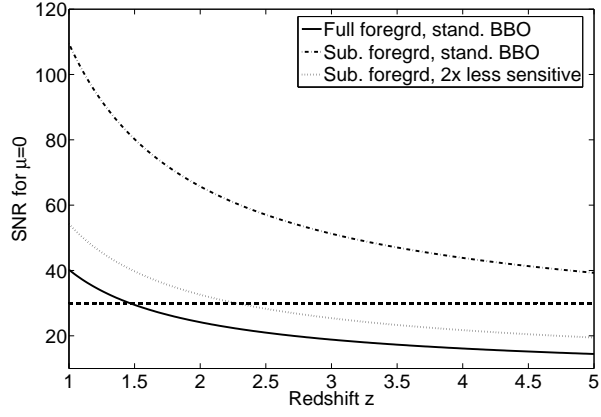}}
\caption[SNR-redshift correspondence]{Shows the SNR ratio of NS-NS mergers with $\mu=0$ in Eq.~(\ref{eqSN}). The total noise 
level is different for each curve. The solid curve is for standard BBO instrumental noise plus confusion noise from {\it all}
NS binaries.  The dotted curve is for ``standard/2'' instrumental noise plus foreground corresponding to
$\dot n_0 = 10^{-7}\,{\rm yr}^{-1}{\rm Mpc}^{-3}$, with $F^\prime_{30} = 0.077$  The highest curve is for standard-BBO instrumental noise and zero
foreground noise. The horizontal line just highlights $\rm SNR = 30$.} 
\label{figShellzSN}
\end{figure}

Here we expand on the results of the previous subsection, to improve understanding. In Fig.~\ref{figShellzSN} 
we plot the SNR of NS binaries
having $\mu =0$ (i.e, those seen edge-on: the least detectable case) as a function of $z$,  under three different assumptions.
The lowest curve (solid line) assumes standard BBO instrumental noise and assumes that the foreground confusion noise is the full $S_{\rm h}^{\rm NSm}(f)$ 
(i.e., the level before any subtraction), with $\dot{n}_0=10^{-7}\,{\rm yr}^{-1}{\rm Mpc}^{-3}$.  In this case, assuming
$\rho_{\rm th} = 30$, all binaries out to $z \approx 1.5$ could be detected, even without first subtracting out the brightest sources.
(And of course, the binaries with more favorable orientations could be detected even farther out.)
In an iterative subtraction scheme, one would begin by subtracting out all the high-SNR sources, which would lower
the total noise and allow one to ``look deeper'' in succeeding iterations. For standard BBO, this iterative scheme reaches the point where
there are {\it zero}, or almost zero,  unsubtracted sources, and then the total noise is just the instrumental noise.  

The SNR for this ``instrumental noise only'' case is shown in the upper (dot-dashed) curve
in Fig.~\ref{figShellzSN}.  
>From this curve one sees immediately that $F=0$ is indeed a self-consistent solution: even the sources with $\mu = 0$ at
$z=5$ are detectable.  What Fig.~\ref{figShellzSN} {\it cannot} show is whether $F=0$ is the {\it only} self-consistent 
solution, but the rest of our analysis shows that this is true (again, for standard BBO sensitivity and $\rho_{\rm th} \le 30$). 
This has the practical implication that our envisioned iterative subtraction procedure should
not get ``stuck'' at some higher $F$ value: it can keep going until all binaries have been removed.
The situation for standard/2 sensitivity is different, as illustrated by the
middle (dotted) curve,  which corresponds to the case $\rho_{\rm th} = 30$ and $\dot n_0 = 10^{-7}\,{\rm yr}^{-1}{\rm Mpc}^{-3}$.
For this case $F^{\prime} = 0.077$, so the unsubtractable foreground noise is small compared to the instrumental noise, and the
SNRs are roughly half the standard-BBO values. But then the $\mu = 0$ binaries can only be detected to $z \approx 2.2$.

\begin{figure}[ht!]
\centerline{\includegraphics[width=9.0cm]{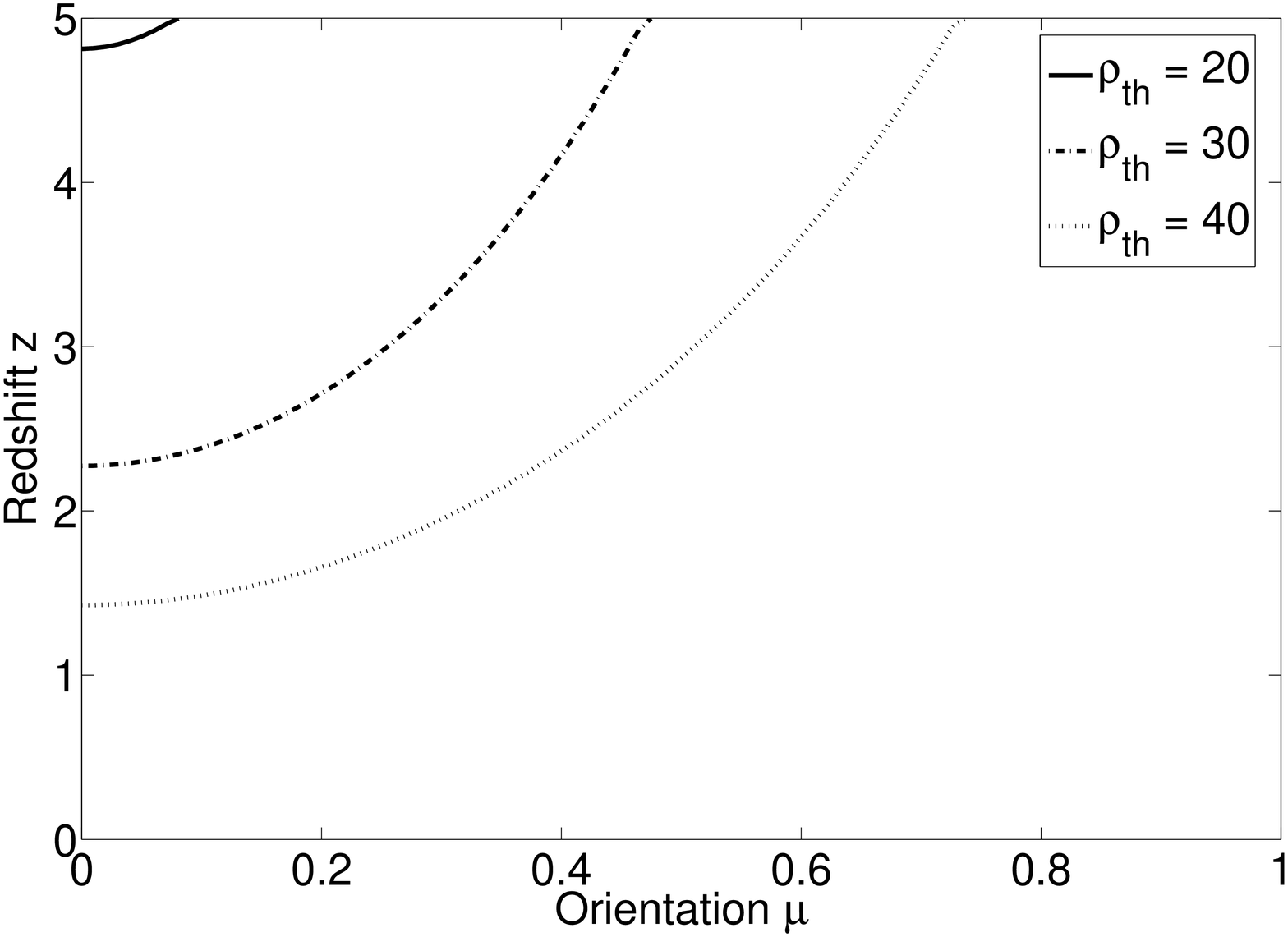}}
\caption[Distance of resolvable NS-NS mergers]{Graph displays the maximum distance to which NS binaries can be detected, as 
function of their orientation angle $\mu \equiv \hat L \cdot \hat N$, for three different detection thresholds $\rho_{\rm th}$.  
Here the instrumental sensitivity is 
``standard/2'' and the merger rate density is $\dot{n}_0=10^{-7}\,{\rm yr}^{-1}{\rm Mpc}^{-3}$ for all three curves.}
\label{figDeclA}
\end{figure}

The distribution of unsubtractable binaries, for BBO with standard/2 sensitivity, is explored further in Fig.~\ref{figDeclA}, 
which shows the maximal redshift to which NS binaries can be detected, as a function of
$\mu$. The three curves are for our three detection thresholds: $\rho_{\rm th} = 20,30,40$; all assume 
$\dot n_0 = 10^{-7}\,{\rm yr}^{-1}{\rm Mpc}^{-3}$. 
For $\rho_{\rm th} =20$ (solid curve), only a tiny corner of the $(z,\mu)$-space contains sources too weak to be detected, and
the number of sources occupying that corner would be of order one (for $\dot n_0 = 10^{-7}$).  
For $\rho_{\rm th} = 30$ or $40$, the ``undetectable regions'' are 
clearly much larger, and contain several percent (or more) of all sources.

\subsection{Confusion noise from a very strong NSm foreground}
The results for $F$ in Eq.~(\ref{eqFBBO}) had basically no 
dependence on the merger rate $\dot n_0$, and the 
$F^{\prime}$  results Table~II showed only weak dependence on
$\dot n_0$, except at the highest values of $\dot n_0$ and $\rho_{\rm th}$. 
The reason for this is simple: for BBO to succeed, the unsubtracted foreground noise must be 
smaller than the primordial background. Therefore, for BBO even to be
``in the right ballpark'', the unsubtracted foreground must be well below the
instrumental noise level.  In this regime, the SNR of
any source is set almost entirely by $S_{\rm h}^{\rm inst}$, and so 
is insensitive to $\dot n_0$. Our results are consistent with the
fact that, even with sensitivity degraded by a factor 2,  
BBO would still be ``in the ballpark'' (albeit insufficient for high $\rho_{\rm th}$).

\begin{figure}[ht]
\centerline{\includegraphics[width=9.0cm]{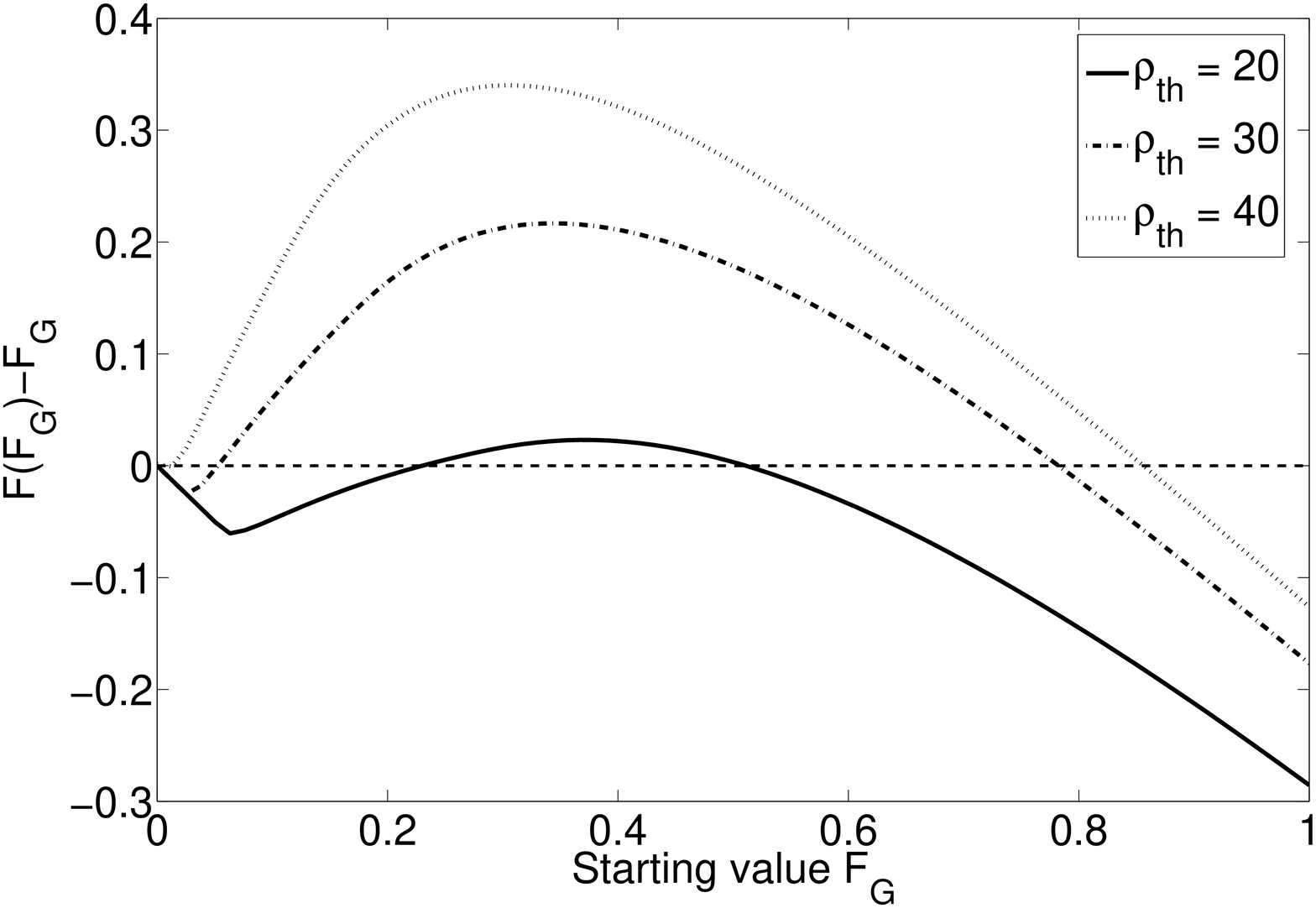}}
\caption[Binary subtraction (increased $\dot{n}_0$)]{Plots $F(F_{\rm G}) - F_{\rm G}$ for an unrealistically high merger rate density, 
$\dot{n}_0=10^{-5}\,{\rm yr}^{-1}{\rm Mpc}^{-3}$. In contrast to cases with lower
$\dot n_0$, each curve now has two zeroes. However only the higher zero (larger $F$ value)
can be reached by an iterative subtraction scheme.}
\label{figFixS4}
\end{figure}

However the dependence of $F$ on $\dot n_0$ becomes
greater as one increases the merger rate, i.e., as unsubtractable binaries come
to represent a significant fraction of the total noise.
Because such cases display the full
utility of our self-consistent method, we here show results for
an unrealistically high merger rate: $\dot{n}_0=10^{-5}\,{\rm yr}^{-1}{\rm Mpc}^{-3}$. 
Fig.~\ref{figFixS4} shows the function $F(F_{\rm
G})-F_{\rm G}$ for this $\dot n_0$, for standard BBO instrumental noise, and
for our $3$ values of $\rho_{\rm th}$.
Interestingly, each curve now has two zeroes; i.e., each case has two self-consistent solutions.  A moment's thought, however, convinces
one that the {\it larger} of the two solutions is the only one that is accessible by an iterative subtraction scheme. 
Such a scheme essentially starts at the
right-most end of the curve and proceeds along it, moving to the left as sources are subtracted, until it reaches the first zero of $F(F_{\rm G}) - F_{\rm G}$.
At that point, any undetected source is too deeply buried in the noise of the other undetected ones (plus the instrumental noise) to be identified.
Otherwise stated: while you can self-consistently ``be at'' the lower-$F_{\rm G}$ solution, the class of schemes we are considering cannot ``bring you there''
(and we suspect that no scheme can).
Thus, we see that in a universe with  $\dot{n}_0=10^{-5}\,{\rm yr}^{-1}{\rm Mpc}^{-3}$, more than half the foreground noise 
would be unresolvable by standard BBO.

\section{Conclusions and Future Work}

We have calculated the efficacy of an iterative procedure for removing
merging NS binary signals from the BBO data stream, as required to
detect any underlying, inflation-generated GW background.
Our calculation basically required as inputs: a) the BBO instrumental noise
curve, b) an estimate of the extragalactic NS-NS merger rate (as a function
of $z$), and c) an estimate of the inspiral SNR required for detection, with
realistic computing power. We find that the current design sensitivity {\it is}
sufficient to allow data analysts to subtract out the merger waveforms, for
the entire range of reasonable merger rates. 
If BBO were less sensitive by a factor 2 (meaning a factor $4$ higher in $S_{\rm h}^{\rm inst}$), 
then BBO's success would depend on having a rather good detection algorithm, capable
of finding almost all sources whose total SNR exceeds $\sim 20$.
If BBO were less sensitive by a factor 4, unsubtractable sources would simply ``cover up''
any underlying primordial background with $\Omega_{\rm GW} \lesssim 10^{-15}$ (or somewhat lower
if NS-NS merger rates are at the low end of the predicted range).

Our goal was to estimate the efficacy of an iterative subtraction procedure,
{\it without actually trying to implement it}.
Of course,
simulations of this procedure, to confirm our calculation or reveal
holes in the argument, would also be very interesting. In particular,
it would be important to confirm that our proposed projection
technique on the cleaned data stream {\it does} sufficiently de-contaminate it
of residuals from imperfect subtractions of
resolved binaries, as we have assumed in this paper.
A careful simulation of the BBO data analysis process would also lead to a
firmer estimate of the threshold SNR $\rho_{th}$ required for merger
detection in practice, as a function of available computing power.

\acknowledgements
We thank S. Phinney, N. Cornish, and N. Seto for helpful discussions regarding BBO's instrumental noise curve, and we 
thank I. Gholami for providing us with results on optimized pulsar search strategies, assuming
$10^{17}$ Flops of computing power. 
Much of this work was carried out while C.C.~was at the Max-Planck-Institut f\"ur Gravitationsphysik (AEI-Golm). C.C.'s  work on this paper was completed at the Jet Propulsion Laboratory, California Institute of Technology, and so was also sponsored by the National Aeronautics and Space Administration.  C.C.~also gratefully acknowledges support from NASA Grant NAG5-12834. 
J.H.~is working for the SFB407/B12 of the Deutsche Forschungsgemeinschaft and J.H.~also thanks 
the Max-Planck-Institut f\"ur Gravitations- physik (AEI-Hannover) for its support. 

\appendix

\section{Derivation of Eq.~(\ref{eqSN})}

In this Appendix we derive Eq.~(\ref{eqSN}).
We begin by averaging over all angles, including $\mu \equiv \hat L \cdot \hat N$; we return to the 
$\mu$-dependence near the end.

Consider first a
single synthetic Michelson data stream from a
single LISA-like detector.
Let the waveform at the detector be $h_{ij}(t) =
h_+(t) e^+_{ij}  + h_\times(t) e^\times_{ij}$, where
$e^+_{ij}$  and $e^\times_{ij}$ are ``+'' and ``$\times$'' polarization tensors,
respectively. The average matched-filter $\rm SNR^2$ for some source
(where the average is over source-direction and polarization angle) is
given by
\be\label{snr2}
<{\rm SNR}^2> = 4 \int_0^\infty
\frac{ |\tilde h_+(f)|^2 + |\tilde h_\times(f)|^2 \drm f}{S_{\rm h}(f)}
\ee
where, as throughout this paper,
$S_{\rm h}(f)$ is the ``sky-averaged'' noise spectral density.
Parseval's Theorem states that
\be
\int_0^\infty |\tilde h_+(f)|^2 \drm f =
\frac{1}{2} \int_{-\infty}^{\infty}  h_+^2(t) \drm t \, ,
\ee
and similarly for $h_\times$
so for a chirping signal with a slowly changing frequency f(t), it is
clear that
\be
|\tilde h_+(f)|^2 + |\tilde h_\times(f)|^2
= \frac{1}{2} \big(\bar h^2_+(t)+
\bar h^2_\times(t)\big) \drm t/\drm f \, ,
\ee
where the overbar denotes time-averaging.

For now, consider some GW source at low redshift ($z << 1$). Then the
rate at which the source loses energy due to GW emission is
\be
\dot E(t) = 4\pi D^2 \, (\pi f^2/4)\, <\, \bar h_+^2(t) + \bar h_\times^2(t) \,>
\ee
where $D$ is its distance, and where the averaging is over all directions from the source.
Therefore we have
\be\label{h2}
<\,|\tilde h_+(f)|^2 + |\tilde h_\times(f)|^2 \, > = \frac{1}{2} \frac{\dot E}{\pi^2 D^2 f^2} \,\drm t/\drm f \, .
\ee

The product $\dot E (\drm t/\drm f)$  equals $|\drm E/\drm f|$.  For a circular-orbit 
binary, the energy  is approximately 
\bea
E &\approx& - \frac{1}{2}\mu M/r \approx- \frac{1}{2}\mu (M \pi f)^{2/3} \label{1st} \\
&= & - \frac{1}{2}{\cal M}^{5/3} (\pi f)^{2/3} \label{2nd}
\eea
from  which we obtain
\be
|\drm E/\drm f| \approx  \frac{1}{3}\,{\cal M}^{5/3}\,\pi^{2/3}f^{-1/3} \, .
\ee
Using this result along with Eqs.~(\ref{snr2}) and (\ref{h2}), we arrive at
\be\label{one-detector}
<\rho^2> = \frac{2 {\cal M}^{5/3}}{3 \pi^{4/3} D^2}
\int_0^\infty{\frac{f^{-7/3}\,\drm f}{S_{\rm h}(f)}} \, .
\ee

The generalization of Eq.~(\ref{one-detector}) to arbitrary redshift is accomplished by the standard replacement~\cite{Markovic}
${\cal M} \rightarrow {\cal M}(1+z)$ and $D \rightarrow D_{\rm L}$, where $D_{\rm L}$ is the luminosity distance.  
The $\mu$-dependence of the waveform's strength--i.e., the $f(\mu)$ factor
in Eq.~(\ref{eqSN})--follows almost immediately from, e.g., Eqs.(2a-2b) of ~\cite{ACST1994}. Finally, to arrive at
Eq.~(\ref{eqSN}), we multiply the rhs of Eq.~(\ref{one-detector}) by a factor of $8$, to account for the fact that
at low-to-mid frequencies BBO is approximately equivalent to $8$ independent Michelson detectors, each with
the same noise density $S_{\rm h}(f)$.

\bibliography{references}

\end{document}